\documentclass[aps,prd,preprint,amsmath,amssymb]{revtex4}
\usepackage{graphicx}
\usepackage{dcolumn}
\usepackage{bm}

\newcommand{\be}{\begin{equation*}}
\newcommand{\ee}{\end{equation*}}
\newcommand{\bea}{\begin{eqnarray}}
\newcommand{\eea}{\end{eqnarray}}
\newcommand{\bean}{\begin{eqnarray*}}
\newcommand{\eean}{\end{eqnarray*}}
\newcommand{\ppb}{{$\bar{p}p \to \bar{p}n \pi^+ $}}
\newcommand{\pp}{{$p p \to p n \pi^+ $}}
\newcommand{\n}{{$N^*$}}
\newcommand{\de}{{$\Delta$}}

\begin{document}

\title{Proposal for Studying $N^*$ Resonances with $\bar{p}p \to \bar{p}n \pi^+ $ Reaction }

\author{
Jia-Jun Wu,$^{1,2}$ Zhen Ouyang$^{2,3}$ and B.~S.~Zou$^{1,2}$\\
$^1$ Institute of High Energy Physics, CAS, P.O.Box 918(4), Beijing 100049, China\\
$^2$ Theoretical Physics Center for Science Facilities, CAS, Beijing 100049, China\\
$^3$ Institute of Modern Physics, CAS, Lanzhou 730000, China}

\date{September 8, 2009}

\begin{abstract}
A theoretical study of the $\bar{p}p \to \bar{p}n \pi^+$ reaction
for antiproton beam energy from 1 to 4 GeV is made by including
contributions from various known $N^*$ and $\Delta^*$ resonances. It
is found that for the beam energy around 1.5 GeV, the contribution
of the Roper resonance $N^*_{(1440)}$ produced by the t-channel
$\sigma$ exchange dominates over all other contributions. Since such
a reaction can be studied in the forthcoming $\bar{P}$ANDA
experiment at the GSI Facility of Antiproton and Ion Research
(FAIR), the reaction will be realistically the cleanest place for
studying the properties of the Roper resonance and the best place
for looking for other ``missing" $N^*$ resonances with large
coupling to $N\sigma$.
\end{abstract}


\maketitle

\section{Introduction}
\label{s1}

The study of $N^*$ resonances can provide us with critical insight
into the nature of QCD in the confinement domain~\cite{ni1}. In the
study of the $N^*$ resonances, there are two long-standing central
issues. First, many $N^*$ states predicted by quark models have not
been observed in experiments~\cite{lbc,cap2,cap3,ni2}, {\sl i.e.},
so-called missing $N^*$ problem. Second, the properties of the
lowest well-established $N^*$ resonances, $N^*_{(1440)}$ and
$N^*_{(1535)}$, are still not well determined
experimentally~\cite{pdg} and not well understood
theoretically~\cite{cap2}.

As the lowest excited nucleon state, the Roper resonance
$N^*_{(1440)}$ was first deduced by $\pi N$ phase shift analysis;
its structure has been arousing people's interests intensely all the
time; {\sl i.e.,} it is lighter than the first odd-parity nucleon
excitation, the $N^*_{(1535)}$, and has a significant branching
ratio into two pions. Up to now, although the existence of the Roper
resonance is well established (four-star ranking in the particle
data book), its properties, such as mass, width, and decay branching
ratios, still suffer large experiment uncertainties~\cite{pdg}.
There are many models on this Roper resonance. In classical quark
models, the Roper resonance has been associated with the first
spin-parity $J^P = 1/2^+$ radial excited state of the
nucleon~\cite{ni2,cap,glo,liu}. In the bag~\cite{mei} and Skyrme
models~\cite{haj}, it was interpreted as surface oscillation, also
called breathing mode. It has also been predicted as a monopole
excitation of the nucleon with the gluonic
excitation~\cite{bar,haqq,kiss} or as dynamically generated from
meson-nucleon interactions~\cite{krehl,sch}. But these predictions
always reach either a larger value for its mass or a much smaller
one for its width and also meet difficulties in explaining its
electromagnetic coupling~\cite{sar}.

Up to now, our knowledge on $N^*$ resonances has been mainly coming
from $\pi N$ and $\gamma N$ experiments. Then those unobserved
missing $N^*$ resonances may be due to their weak couplings to $\pi
N$ and $\gamma N$. Even for the well-established Roper resonance,
its properties can be extracted only by detailed partial-wave
analysis. No corresponding peak has been observed from the $\pi N$
invariant mass spectrum because of its nearby strong $\Delta$ peak.
A difficulty in extracting the $N^*$ information from these
experiments is the isospin decomposition of $1/2$ and
$3/2$~\cite{work}. Recently, the $J/\psi \to \bar{N}N\pi$ and $pp
\to pn\pi^+$ reactions have been used to study $N^*$ resonances with
claimed observation of the Roper resonance peak~\cite{abli,clement}
due to their isospin filter effect~\cite{ouyang1,ouyangxie}.
However, because of the presence of large interfering contributions
from other resonances, there is still considerable model dependence
in extracting its properties. Moreover, the data~\cite{clement} from
$pp \to pn\pi^+$ reaction are based on a preliminary analysis of the
limited phase space, suffering a strong model dependence, and may
have changed quite a bit in the course of the analysis, as suggested
in a more recent paper by the same collaboration~\cite{Skorodko}.

In this work, we propose to study the Roper and other $N^*$
resonances with the $\bar{p}p \to \bar{p}n\pi^+$ reaction, where
thanks to the absence of the $\Delta^{++}$ state, the contribution
of the $\Delta$ excitation is much smaller than that in the $pp \to
pn\pi^+$ reaction. It is found that for the beam energy around 1.5
GeV, the contribution of the Roper resonance $N^*_{(1440)}$ produced
by the t-channel $\sigma$ exchange dominates over all other
contributions because of its known large coupling to
$N\sigma$~\cite{pdg,hir}. This will provide the cleanest place for
studying the properties of the Roper resonance and the best place
for looking for other missing $N^*$ resonances with large coupling
to $N\sigma$.

Such a reaction can be studied by the scheduled experiments on the
Proton Antiproton Detector Array (PANDA) at the GSI Facility for
Antiproton and Ion Research (FAIR) with the antiproton beam of
kinetic energy ranging from 1 to 15 GeV~\cite{pan1}. The detector
with an almost $4\pi$ detection coverage for both charged particles
and photons can detect $\pi^+$ and $\bar p$ in the final state. The
neutron can be reconstructed from a missing mass spectrum against
the $\pi^+$ and $\bar p$. Hence we suggest the PANDA Collaboration
pay good attention to the study of $N^*$ resonances, considering its
unique advantages found in this work.

In the next section, we present the formalism and ingredients for
the calculation of the $\bar{p}p \to \bar{p}n\pi^+$ reaction by
including various intermediate \n\ and $\Delta^*$ resonances. Then
in the Sec.\ref{s3} we give the numerical results of the
calculation, compare this reaction to the $pp \to p n\pi^+$
reaction, and discuss the results.

\section{Formalism and ingredients}
\label{s2}

We study the \ppb\ reaction within an effective Lagrangian approach.
All the basic Feynman diagrams involved in our calculation for this
reaction are depicted in Fig. 1. The formalism and ingredients are
very similar to those used in the study of the \pp\
reaction~\cite{ouyangxie}, where only $N^*_{(1440)}$,
$N^*_{(1520)}$, $N^*_{(1680)}$ and $\Delta_{(1232)}$ resonances are
found to play significant roles for the beam energy around
$T_p=1\sim 3$ GeV. With the experience on the \pp reaction, we
investigate here the contribution from these resonances to the
present \ppb reaction for the beam energy $T_{\bar p}=1\sim 4$ GeV.

\begin{figure}[htbp] \vspace{-0.cm}
\begin{center}
\includegraphics[width=0.9\columnwidth]{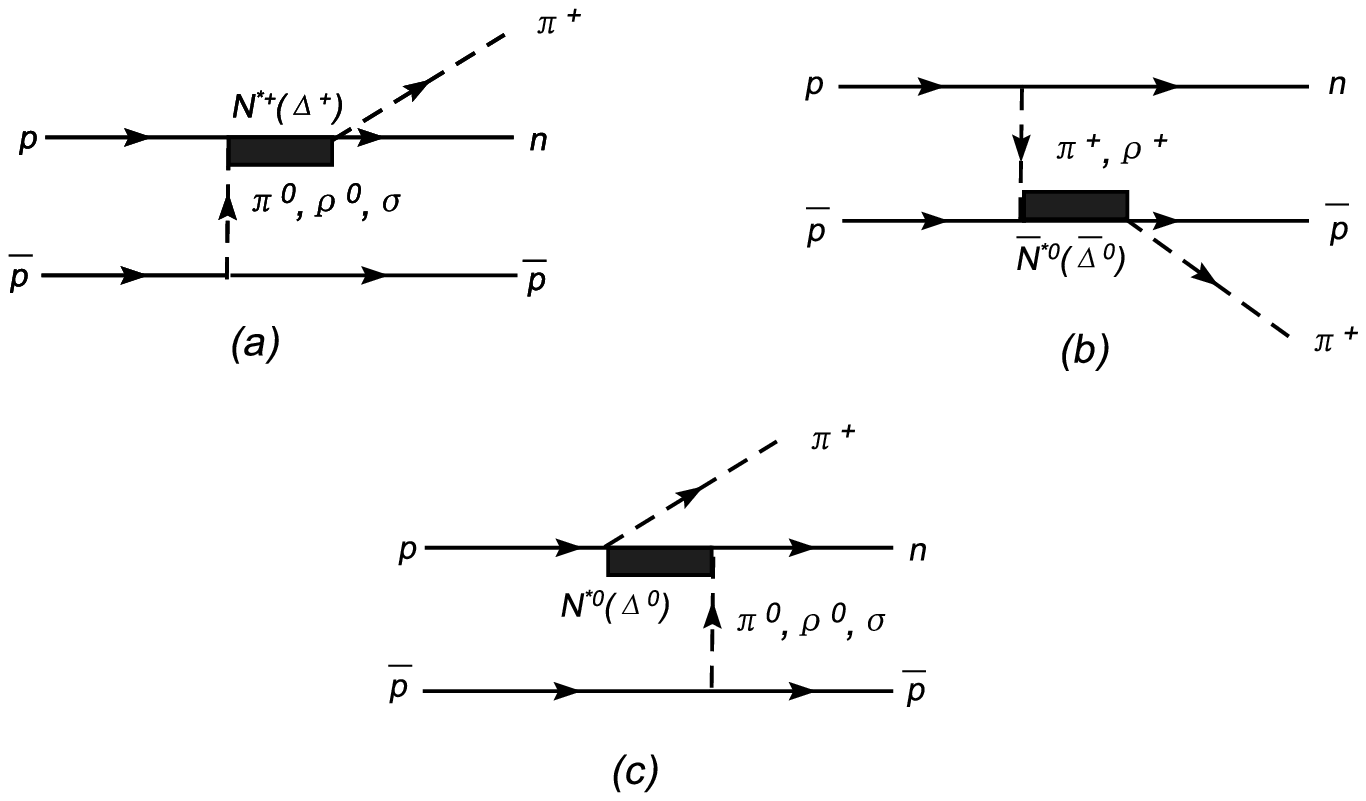}
\caption{Feynman diagrams of the \ppb reaction. } \label{f1}
\end{center}
\end{figure}

First, we give the effective Lagrangian densities for describing the
meson-$NN$ vertices:
\begin{eqnarray}
{\cal L}_{\pi NN}&=&g_{\pi NN}\bar{u}_{N}\gamma_{5}\vec{\tau} \cdot
\vec{\psi}_{\pi}
u_{N}+h.c.,\\
{\cal L}_{\sigma NN}&=&g_{\sigma NN}\bar{u}_{N} \psi_{\sigma} u_{N}+h.c., \\
{\cal L}_{\rho NN}&=&g_{\rho
NN}\bar{u}_{N}\left(\gamma_{\mu}+\frac{\kappa}{2m_{N}}\sigma_{\mu\nu}\partial^{\nu}\right)\vec{\tau}
\cdot \vec{\psi}_{\rho} u_{N}+h.c.. \label{mnn}
\end{eqnarray}
Here $\vec{\tau}$ is the usual isospin-$1/2$ Pauli matrix operator,
and the coupling constants are all listed in Table~\ref{tab1}. At
each vertex, we need a relevant off-shell form factor for the
exchanged meson. In this paper, we use the same form factors as
assumed in the previous
literature~\cite{xie1,ouyang1,ouyangxie,kt,as,holinde,rm}:
\begin{eqnarray}
F^{NN}_{M}(k_{M}^{2})=\left(\frac{\Lambda_{M}^{2}-m_{M}^{2}}{\Lambda_{M}^{2}-k_{M}^{2}}\right)^{n}.\label{bonn}
\end{eqnarray}
Here M represents $\pi$, $\sigma$, or $\rho$ mesons. The $\Lambda_M$
parameters as used in Refs.\cite{ouyang1,ouyangxie} for the $pp\to
pn\pi^+$ reaction are also listed in Table~\ref{tab1}. Note that for
$NN$ elastic scattering, the square of the four-momentum vector
$k_M$ is equal to its corresponding three-momentum squared with a
minus sign; hence, in some literature, such as in the Bonn
model~\cite{rm}, an equivalent formula of the form factor with the
three-momentum is used.

\begin{table}[ht]
\begin{tabular}{c c c c}
\hline\hline\  $M$\ \  &\  $n$\ \ & $g^2_{MNN}/4\pi$ & $\Lambda_{M}(GeV)$ \\
\hline
 $\pi$ & 1 &\  14.4\  & 1.3 \\
 $\sigma$ &\  1\  & 5.69 & 2.0 \\
 $\rho$ & 2 &\  0.9 ($\kappa=6.1$)\  & 1.85 \\
\hline \hline
\end{tabular}
\caption{Coupling constants and cutoff parameters used for the
meson-$NN$ vertices~\cite{ouyang1,ouyangxie}.} \label{tab1}
\end{table}

Second, we consider the interaction vertices involving \n\ and
$\Delta^*$ resonances. In Ref.~\cite{zouf}, a Lorentz covariant
orbital-spin scheme for $N^*NM$ couplings is described in detail and
can be easily extended to describe all the couplings appearing in
the Feynman diagrams in Fig.~\ref{f1}. By using that scheme, the
relevant important effective couplings~\cite{ouyang1,ouyangxie} are:
\begin{eqnarray}
{\cal L}_{\pi N\Delta_{(1232)}}&=&g_{\pi
N\Delta_{(1232)}}\bar{u}_{N}
\partial^{\mu}\psi_{\pi}\tilde{\tau}
u_{\Delta_{(1232)}\mu}+h.c.,\\
{\cal L}_{\sigma NN^*_{(1440)}}&=&g_{\sigma
NN^*_{(1440)}}\bar{u}_{N} \psi_{\sigma} u_{N^*_{(1440)}}+h.c., \\
{\cal L}_{\pi NN^*_{(1440)}}&=&g_{\pi
NN^*_{(1440)}}\bar{u}_{N}\gamma_{5}\gamma_{\mu}\vec{\tau}
\cdot\partial^{\mu} \vec{\psi}_{\pi} u_{N^*_{(1440)}}+h.c., \\
{\cal L}_{\pi NN^*_{(1520)}}&=&g_{\pi
NN^*_{(1520)}}\bar{u}_{N}\gamma_{5}\gamma_{\mu}p_{\pi}^{\mu}p_{\pi}^{\nu}\vec{\tau}
\cdot \vec{\psi}_{\pi} u_{N^*_{(1520)}\nu}+h.c., \\
{\cal L}_{\rho NN^*_{(1520)}}&=&g_{\rho
NN^*_{(1520)}}\bar{u}_{N}\vec{\tau} \cdot
\vec{\psi}^{\mu}_{\rho} u_{N^*_{(1520)}\mu}+h.c.,\\
{\cal L}_{\pi NN^*_{(1680)}}&=&g_{\pi
NN^*_{(1680)}}\bar{u}_{N}\gamma_{5}\gamma_{\mu}p_{\pi}^{\mu}p_{\pi}^{\nu}p_{\pi}^{\lambda}
\vec{\tau} \cdot \vec{\psi}_{\pi} u_{N^*_{(1680)}\nu\lambda}+h.c. .
\label{res}
\end{eqnarray}
Here $\tilde{\tau}$ is the $\frac{1}{2} \leftrightarrow \frac{3}{2}$
isospin transition operator. For the t-channel exchanged meson
attached to every \n \ and \de \ resonance, we also need the
off-shell form factor:
\begin{eqnarray}
F^{NR}_{M}(k_{M}^{2})=\left(\frac{(\Lambda^R_{M})^{2}-m_{M}^{2}}{(\Lambda^R_{M})^{2}-k_{M}^{2}}\right)^n.\label{refa}
\end{eqnarray}
where $R$ is \n \ or \de. For the s-channel baryon resonances in
Figs. 1(a) and 1(b) or u-channel baryon resonances in Fig. 1(c) we
use the off-shell form factor~\cite{gp,vs,tfe}
\begin{eqnarray}
F_R(q^2)=\frac{\Lambda^{4}}{\Lambda^{4}+(q^{2}-m^{2}_{R})^{2}}\label{ba}
\end{eqnarray}
with $\Lambda=0.8GeV$.

Although only the resonances and the meson exchanges listed in
Table~\ref{tab2} are included in our present calculation, the
results will not change much if all other \n and $\Delta^*$
resonances with spin-parity $1/2^\pm$, $3/2^\pm$ and $5/2^\pm$
listed in the PDG~\cite{pdg} or other meson exchanges are also
included, according to results from Ref.~\cite{ouyangxie} for the
\pp\ reaction.

The coupling constants of resonances can be obtained from their
experimentally observed partial decay widths. For example, the
$g_{N^*_{(1440)}N\pi^{0}}$ can be obtained by the formula:
\begin{eqnarray}
\Gamma_{N^*_{(1440)} \rightarrow
N\pi}&=&\frac{g^{2}_{N^*_{(1440)}N\pi^{0}}p^{c.m.}_{N}}{4\pi}
\left[\frac{m_{\pi}^{2}(E_{N}-m_{N})}{m_{N^*_{(1440)}}}+2(p^{c.m.}_{N})^{2}\right],
\end{eqnarray}
with
\begin{eqnarray}
p^{c.m.}_{N}&=&\sqrt{\frac{(m^{2}_{N^{*}_{(1440)}}-(m_{N}+m_{\pi})^{2})(m^{2}_{N^{*}_{(1440)}}
-(m_{N}-m_{\pi})^{2})}{4m^{2}_{N^{*}_{(1440)}}}},\\
E_{N}&=&\sqrt{(p^{c.m.}_{N})^{2}+m^{2}_{N}}.
\end{eqnarray}
In Table \ref{tab2}, we list all the coupling constants and
$\Lambda^{R}_{M}$ parameters used in the calculation.

Third, we give the propagators of relevant particles. For the $\pi$,
$\sigma$, and $\rho$ mesons, their propagators are simple:
\begin{eqnarray}
G_{\pi(q)}&=&\frac{1}{q^{2}-m_{\pi}^{2}},\\
G_{\sigma(q)}&=&\frac{1}{q^{2}-m_{\sigma}^{2}},\\
G_{\rho(q)}&=&\frac{-\tilde{g}_{\mu\nu}}{q^{2}-m_{\rho}^{2}}.\label{bp1}
\end{eqnarray}
For the \n\ and \de\  resonances, they are spin-1/2, spin-3/2, and
spin-5/2 resonances. In addition, we must consider their
antiparticles. The general formulas for the propagator of a
half-integral spin particle is~\cite{ruan,reb}
\begin{eqnarray}
G^{n+\frac{1}{2}(\pm)}_{R(q)}&=&\frac{P^{n+\frac{1}{2}(\pm)}_{\mu_{1}\mu_{2}...\mu_{n}\nu_{1}\nu_{2}...\nu_{n}}}
{q^{2}-m_{R}^{2}+im_{R}\Gamma_{R}},\\
P^{n+\frac{1}{2}(\pm)}_{\mu_{1}\mu_{2}...\mu_{n}\nu_{1}\nu_{2}...\nu_{n}}&=&\frac{n+1}{2n+3}
(\not\! p
 \pm m)\gamma^{\alpha}\gamma^{\beta}P^{n+1}_{\alpha\mu_{1}\mu_{2}...\mu_{n}\beta\nu_{1}\nu_{2}...\nu_{n}},\\
P^{n}_{\mu_{1}\mu_{2}...\mu_{n}\nu_{1}\nu_{2}...\nu_{n}}&=&\left(\frac{1}{n!}\right)^2\sum_{P_{(\mu)}P_{(\nu)}}
\big[\prod_{i=1}^{n}
P_{\mu_{i}\nu_{i}}+a_{1}P_{\mu_{1}\mu_{2}}P_{\nu_{1}\nu_{2}}\prod_{i=3}^{n}
P_{\mu_{i}\nu_{i}}+...\nonumber\\
&&+a_{r}P_{\mu_{1}\mu_{2}}P_{\nu_{1}\nu_{2}}P_{\mu_{3}\mu_{4}}
P_{\nu_{3}\nu_{4}}...P_{\mu_{2r-1}\mu_{2r}}P_{\nu_{2r-1}\nu_{2r}}\prod_{i=2r+1}^{n}
P_{\mu_{i}\nu_{i}}+...\nonumber\\
&&+\{^{a_{n/2}P_{\mu_{1}\mu_{2}}P_{\nu_{1}\nu_{2}}...P_{\mu_{n-1}\mu_{n}}P_{\nu_{n-1}\nu_{n}}(for\
even\
n)}_{a_{(n-1)/2}P_{\mu_{1}\mu_{2}}P_{\nu_{1}\nu_{2}}...P_{\mu_{n-2}\mu_{n-1}}P_{\nu_{n-2}\nu_{n-1}}(for\
odd\ n)}\big], \\
a_{r(n)}&=&\left(-\frac{1}{2}\right)^{r}\frac{n!}{r!(n-2r)!(2n-1)(2n-3)...(2n-2r+1)}.\label{bp2}
\end{eqnarray}
From these formulas, the propagators of the relevant half-integral
spin particles can be obtained explicitly as follows:
\begin{eqnarray}
G^{\frac{1}{2}(\pm)}_{R(q)}&=&\frac{(\not\! p \pm m)}{q^{2}-m_{R}^{2}+im_{R}\Gamma_{R}},\\
G^{\frac{3}{2}(\pm)}_{R(q)}&=&\frac{(\not\! p \pm
m)}{q^{2}-m_{R}^{2}
+im_{R}\Gamma_{R}}\left(-g_{\mu\nu}+\frac{1}{3}\gamma_{\mu}\gamma_{\nu}
+ \frac{2}{3}\frac{q_{\mu}q_{\nu}}{q^{2}}\pm\frac{1}{3m_{R}}(\gamma_{\mu}q_{\nu}-\gamma_{\nu}q_{\mu})\right),\\
G^{\frac{5}{2}(\pm)}_{R(q)}&=&\frac{(\not\! p \pm
m)}{q^{2}-m_{R}^{2}+im_{R}\Gamma_{R}}[\frac{1}{2}(\tilde{g}_{\mu_{1}\nu_{1}}\tilde{g}_{\mu_{2}\nu_{2}}
+\tilde{g}_{\mu_{1}\nu_{2}}\tilde{g}_{\mu_{2}\nu_{1}})
-\frac{1}{5}\tilde{g}_{\mu_{1}\mu_{2}}\tilde{g}_{\nu_{1}\nu_{2}}\nonumber\\
&&-\frac{1}{10}(\tilde{\gamma}_{\mu_{1}}\tilde{\gamma}_{\nu_{1}}\tilde{g}_{\mu_{2}\nu_{2}}
+\tilde{\gamma}_{\mu_{1}}\tilde{\gamma}_{\nu_{2}}\tilde{g}_{\mu_{2}\nu_{1}}
+\tilde{\gamma}_{\mu_{2}}\tilde{\gamma}_{\nu_{1}}\tilde{g}_{\mu_{1}\nu_{2}}
+\tilde{\gamma}_{\mu_{2}}\tilde{\gamma}_{\nu_{2}}\tilde{g}_{\mu_{1}\nu_{1}})],\\
\tilde{\gamma}_{\nu}&=&\gamma_{\nu}-\frac{q_{\nu}\not\!q}{q^{2}},\
\tilde{g}_{\mu\nu}=g_{\mu\nu}-\frac{q_{\mu}q_{\nu}}{q^{2}}.\label{bp3}
\end{eqnarray}
Here $\pm$ means particle and antiparticle, respectively. We list
the values for the widths $(\Gamma_{R})$ and branching ratios of the
included \n\ and \de\ resonances in Table~\ref{tab2}.

\begin{table}[ht]
\begin{tabular}{ c c c c c c c }
\hline \hline $R$\ \    & \ $n$\ \ \ & $\Gamma_{R}(GeV)$\ & Decay
mode\ &  Branching ratios\
& $g^{2}/4\pi$ & $\Lambda^R_{M}(GeV)$ \\
\hline
 $\Delta_{(1232)}$ & 2        & 0.118            & $N\pi$       & 1.0                  & 19.54       & 0.6 \\
 $N^*_{(1440)}$    & 1        & 0.3              & $N\pi$       & 0.65                 & 1.53        & 1.3 \\
                   &          &                  & $N\sigma$    & 0.075                & 3.20        & 1.1 \\
 $N^*_{(1520)}$    & 1        & 0.115            & $N\pi$       & 0.6                  & 5.19        & 0.8 \\
                   &          &                  & $N\rho$      & 0.09                 & 3.96        & 0.8 \\
 $N^*_{(1680)}$    & 1        & 0.13             & $N\pi$       & 0.675                & 16.59        & 0.8 \\
\hline \hline
\end{tabular}
\caption{Resonances and parameters used in the calculation. Widths
and branching ratios are from PDG~\cite{pdg}; cutoff parameters are
from Refs.~\cite{xie1,ouyang1,ouyangxie,kt}.} \label{tab2}
\end{table}

With all relevant effective Lagrangians, coupling constants, and
propagators fixed, the amplitudes for various diagrams can be
written straightforwardly by following the Feynman rules. And the
total amplitude is just their simple sum. Here we give explicitly
the individual amplitudes corresponding to $N^{*+}_{(1440)}\pi^{0}$,
$N^{*0}_{(1440)}\pi^{0}$ and $\bar{N}^{*0}_{(1440)}\pi^{+}$ for the
Feynman diagrams (a), (c), and (b) in Fig.\ref{f1}, as an example,
\begin{eqnarray}
{\cal M}_{N^{*}_{(1440)}\pi^{0}}&=&{\cal
M}_{N^{*+}_{(1440)}\pi^{0}}+{\cal
M}_{N^{*0}_{(1440)}\pi^{0}}\nonumber\\
&=&\!\frac{\sqrt{2}}{3}\left(\bar{u}_{p_{n}s_{n}}\!\gamma_{5}\!\!\not\!p_{\pi^+}
G^{(\frac{1}{2})+}_{N^*_{(1440)}}\!F_{N^{*}_{(1440)}}\!(q^2)\!\!\not\!k_{\pi^0}u_{p_{p}s_{p}}
\!+\!\bar{u}_{p_{n}s_{n}}\!\!\not\!k_{\pi^0}G^{(\frac{1}{2})+}_{N^*_{(1440)}}
\!F_{N^{*}_{1440}}\!(q^2)\!\!\not\!p_{\pi^+}u_{p_{p}s_{p}}\right)\nonumber\\
&& \times g^2_{\pi
NN^*_{(1440)}}\frac{1}{k^{2}_{\pi^0}-m^{2}_{\pi^{0}}}
F^{NN^{*}}_\pi(k^2_{\pi^0})F^{NN}_{\pi}(k^2_{\pi^0})g_{NN\pi}\bar{v}_{p_{\bar{p}_{1}}s_{\bar{p}_{1}}}
\gamma_{5}v_{p_{\bar{p}_{2}}s_{\bar{p}_{2}}},\\
{\cal M}_{\bar{N}^*_{(1440)}\pi^{+}}&=&\frac{2\sqrt{2}}{3}g^2_{\pi
NN^*_{(1440)}}\bar{v}_{p_{\bar{p}_{1}}s_{\bar{p}_{1}}}\gamma_{5}\not\!p_{\pi^+}
G^{(\frac{1}{2})-}_{\bar{N}^*_{(1440)}}F_{N^{*}_{(1440)}}(q^2)\not\!k_{\pi^+}
v_{p_{\bar{p}_{2}}s_{\bar{p}_{2}}}\frac{1}{k^{2}_{\pi^+}-m^{2}_{\pi^{+}}}\nonumber\\
&&\times
F^{NN^{*}}_\pi(k^2_{\pi^+})F^{NN}_{\pi}(k^2_{\pi^+})g_{NN\pi}\bar{u}_{p_{n}s_{n}}
\gamma_{5}u_{p_{p}s_{p}},\label{M1440}
\end{eqnarray}
where $u_{p_{n}s_{n}}$, $v_{p_{\bar{p}_{2}}s_{\bar{p}_{2}}}$,
$u_{p_{p}s_{p}}$, and $v_{p_{\bar{p}_{1}}s_{\bar{p}_{1}}}$ denote
the spin wave functions of the outgoing neutron, antiproton in the
final state and initial proton and antiproton, respectively.
$p_{\pi^+}$, $k_{\pi^+}$, and $k_{\pi^0}$ are the four-momenta of
the outgoing and the exchanged pion mesons. q is the four-momenta of
the \n. $p_{p}$ and $p_{\bar{p}_{1}}$ represent the four-momenta of
the initial proton and antiproton. $p_{n}$ and $p_{\bar{p}_{2}}$
represent the four-momenta of the final neutron and antiproton. And
factor $\sqrt{2}/3$ and $2\sqrt{2}/3$ are from isospin $C-G$
coefficients.

So the total amplitude of the \ppb\ reaction can be obtained as
\begin{eqnarray}
{\cal M}_{\bar{p}p \to \bar{p}n\pi^{+}}&=&{\cal M}_{p \pi^{0}}+{\cal
M}_{\bar{n} \pi^{+}}+{\cal M}_{N^*_{(1440)}\pi^{0}}+{\cal
M}_{\bar{N}^*_{(1440)}\pi^{+}}+{\cal
M}_{N^*_{(1440)}\sigma}\nonumber\\
&&+{\cal M}_{N^*_{(1520)}\pi^{0}}+{\cal
M}_{\bar{N}^*_{(1520)}\pi^{+}}+{\cal M}_{N^*_{(1520)}\rho^{0}}+{\cal
M}_{\bar{N}^*_{(1520)}\rho^{+}}\nonumber\\
&&+{\cal \ M}_{N^*_{(1680)}\pi^{0}}+{\cal
M}_{\bar{N}^*_{(1680)}\pi^{+}}+{\cal
M}_{\Delta_{(1232)}\pi^{0}}+{\cal
M}_{\bar{\Delta}^{0}_{(1232)}\pi^{+}}.\label{M}
\end{eqnarray}

Then the calculation of the cross section $\sigma_{\bar{p}p
\rightarrow \bar{p}n\pi^{+}}$ is straightforward:
\begin{eqnarray}
\sigma_{\bar{p}p \to \bar{p}n\pi^{+}}&=&
\frac{1}{4}\frac{m^{2}_{p}}{(2\pi)^{5}\sqrt{(p_{p}\cdot
p_{\bar{p}_{1}})^{2}-m^{2}_{p}}} \sum_{s_{i}}\sum_{s_{f}}|{\cal
M}_{\bar{p}p
\to \bar{p}n\pi^{+}}|^{2}d\phi,\\
d\phi&=&\frac{m_{p}d^{3}p_{\bar{p}_{2}}}{E_{\bar{p}_{2}}}\frac{d^{3}p_{\pi}}{2E_{\pi}}
\frac{m_{n}d^{3}p_{n}}{E_{n}}\delta^{4}(p_{p}+p_{\bar{p}_{1}}-p_{n}-p_{\pi}-p_{\bar{p}_{2}}).\label{corss}
\end{eqnarray}

\section{Numerical results and discussion}
\label{s3}

\begin{figure}[htbp] \vspace{-0.cm}
\begin{center}
\includegraphics[width=0.49\columnwidth]{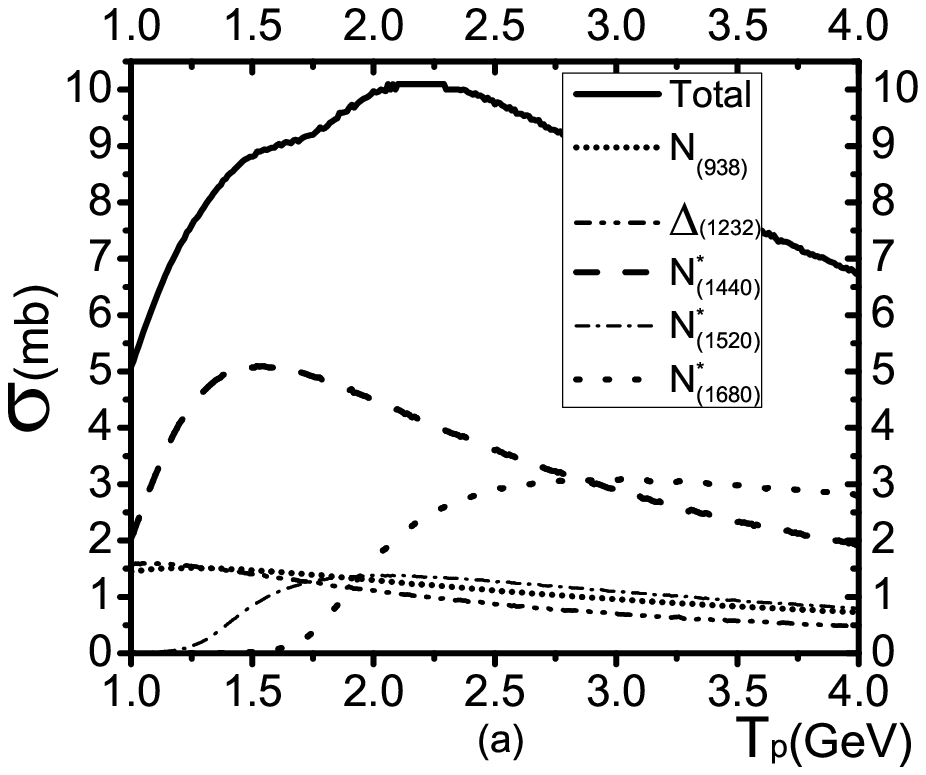}
\includegraphics[width=0.49\columnwidth]{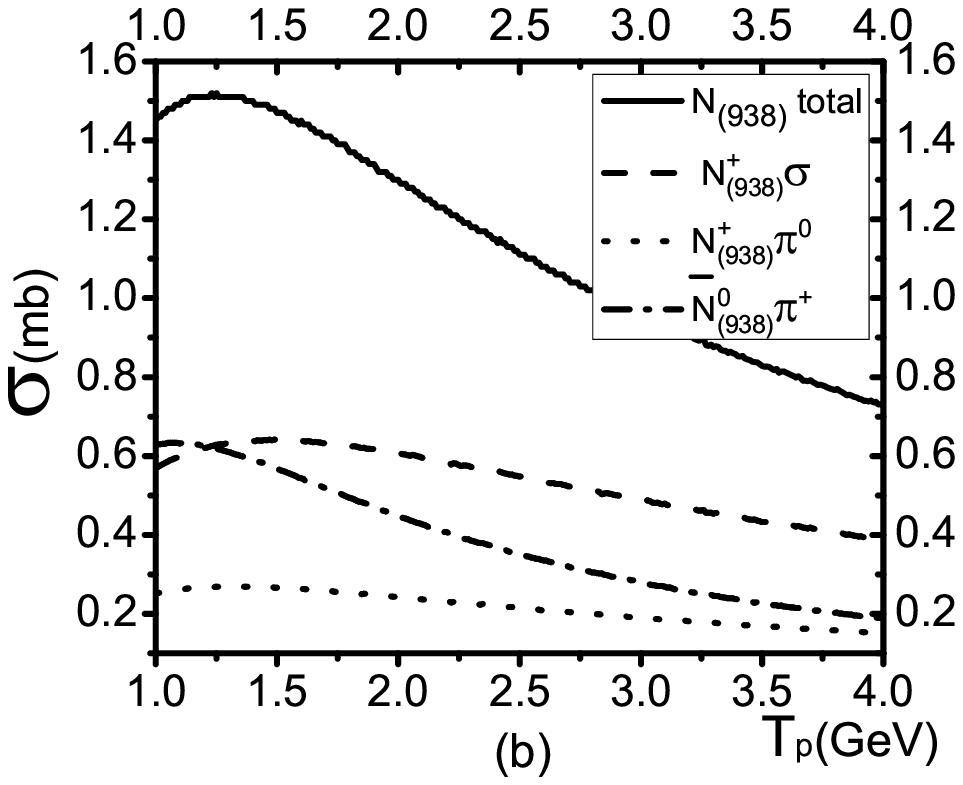}
\includegraphics[width=0.49\columnwidth]{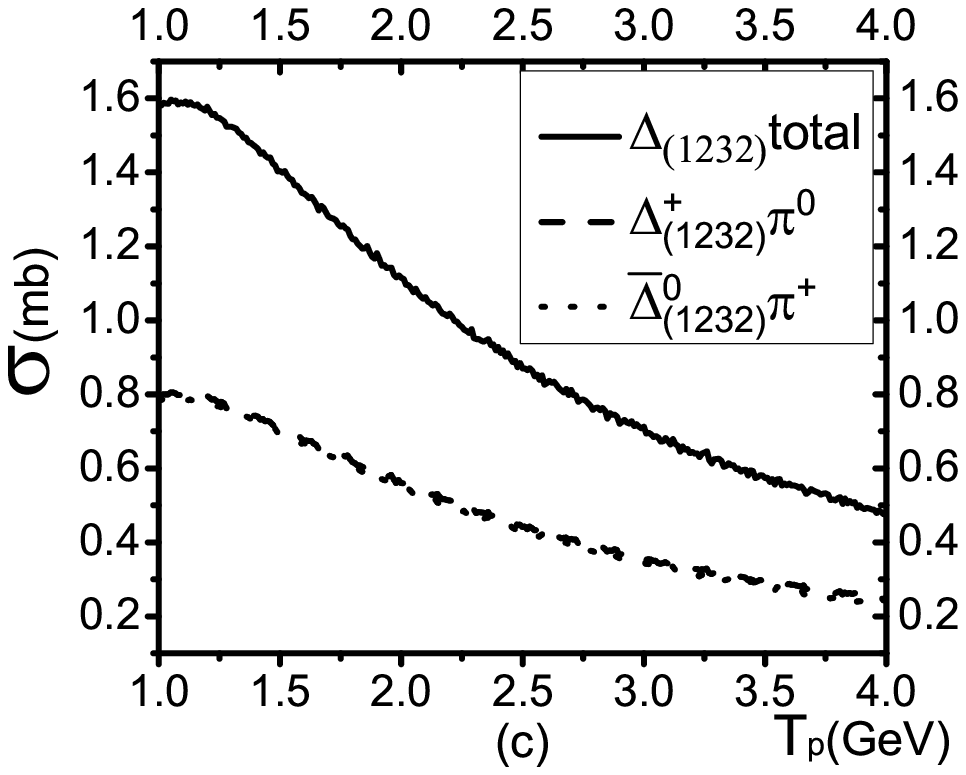}
\includegraphics[width=0.49\columnwidth]{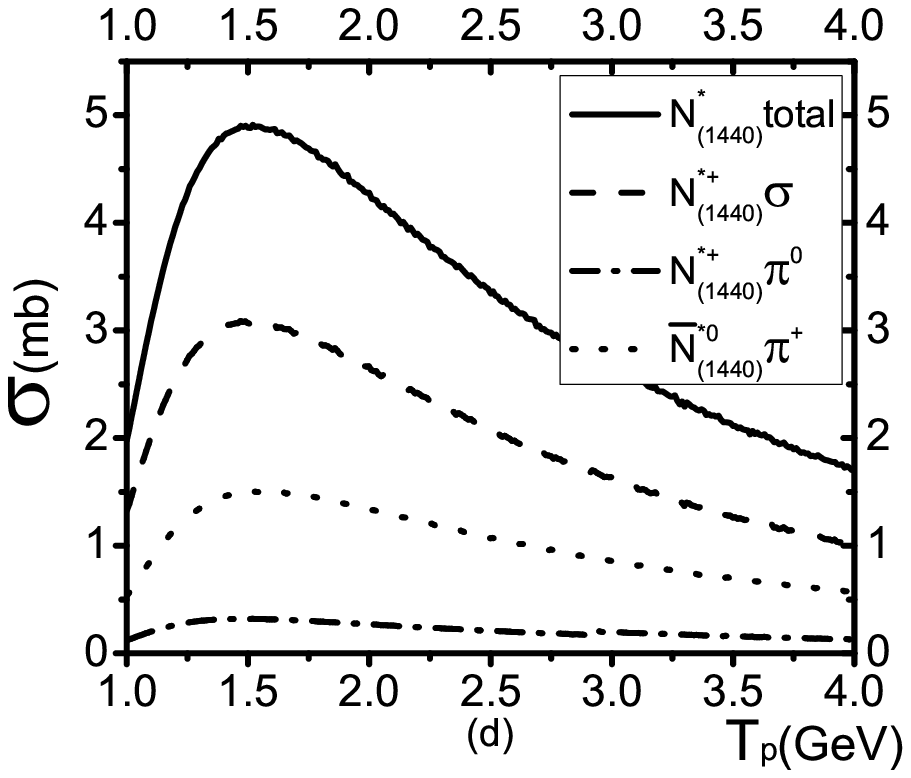}
\includegraphics[width=0.49\columnwidth]{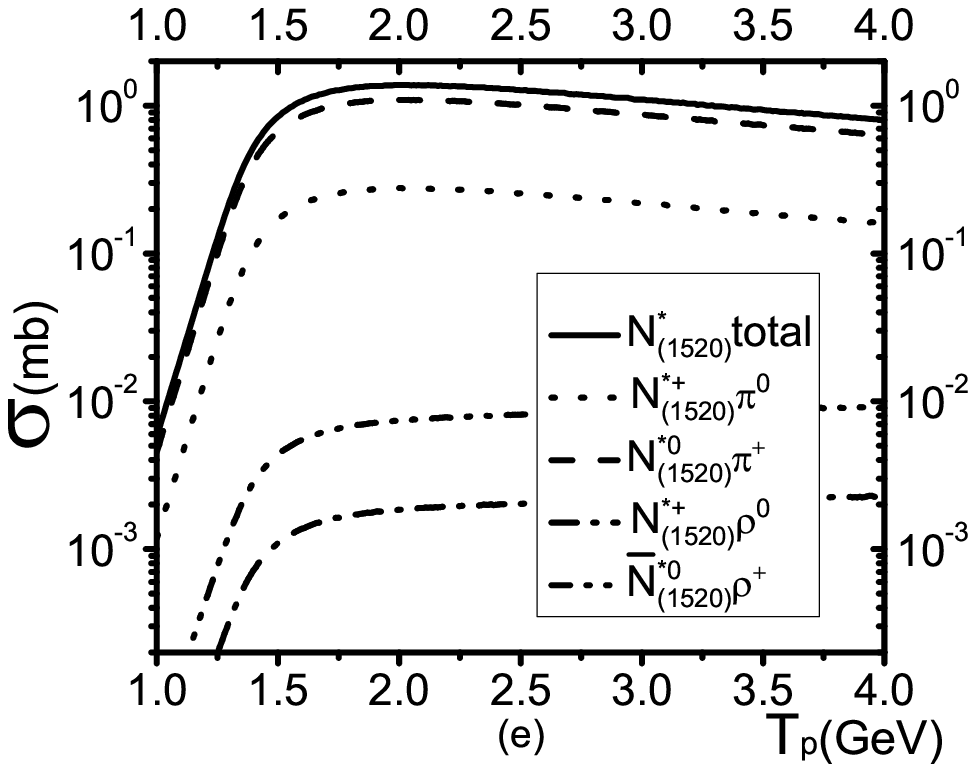}
\includegraphics[width=0.49\columnwidth]{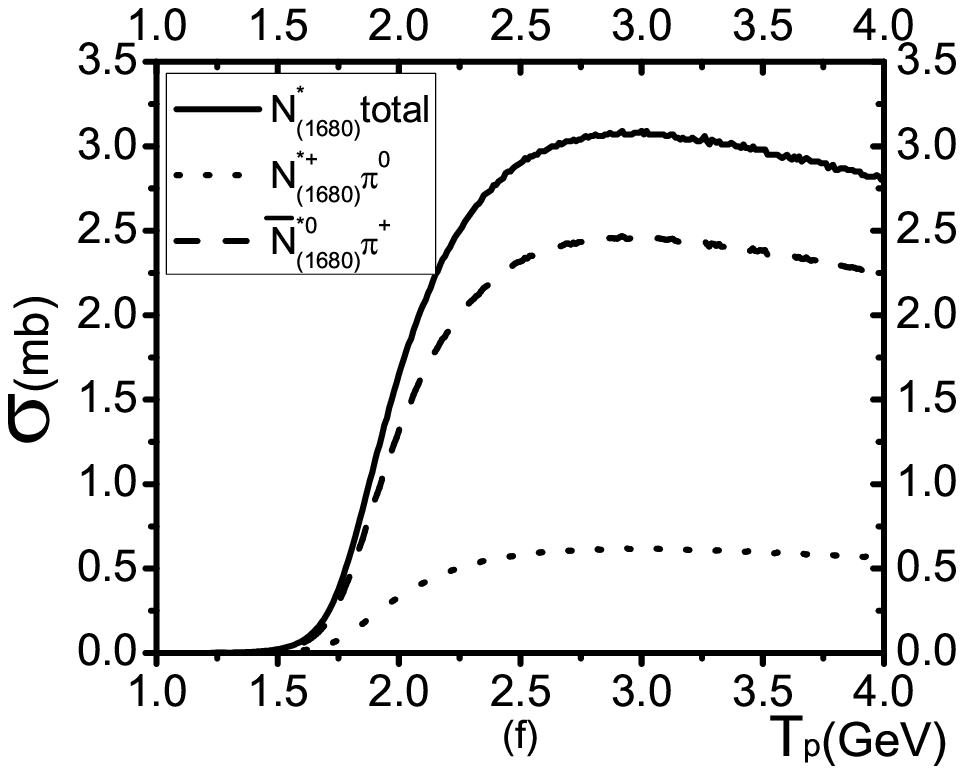}
\caption{Prediction of the cross section vs beam energy $T_{\bar p}$
for the $\bar{p} p \to\bar{ p} n \pi^+$ reaction. (a) Total cross
section and contributions from each resonance included. (b)-(f)
$N_{(938)}$, $\Delta_{(1232)}$, $N^*_{(1440)}$, $N^*_{(1520)}$, and
$N^*_{(1680)}$, respectively, showing contributions from various
Feynman diagrams for each resonance and their subtotal cross
section.} \label{f2}
\end{center}
\end{figure}

With the formalism and ingredients given in the former section, we
compute the total cross section versus the kinetic energy of the
antiproton beam $T_{\bar p}$ for the $\bar{p}p \to \bar{p}n\pi^+$
reaction for $T_{\bar p}=1\sim 4$ GeV by using the code FOWL from
the CERN program library, which is a program for Monte Carlo
multiparticle phase-space integration weighted by the amplitude
squared. The results are shown in Fig.\ref{f2}. The total cross
section for the \ppb \ reaction reaches a maximum of about 10 mb at
$T_{\bar p}$ around 2.2 GeV. Compared with the $\bar pp$ total cross
section of about 90 mb and $\bar pp$ elastic scattering cross
section of about 30 mb around such energy~\cite{pdg}, this is a
rather large share of the $\bar pp$ total cross section.

For the energies from 1 to 2.8 GeV, the largest contribution comes
from the Roper $N^*_{(1440)}$ excitation. It reaches maximum around
1.55 GeV, where it dominates over all other contributions. It is
mainly produced by the t-channel $\sigma$ exchange as shown in Fig.
\ref{f2}(d). This will provide a very clean place for studying
properties of the Roper resonance, such as its mass, width, and
coupling to $N\sigma$. The t-channel $\sigma$ exchange is not only
important for $N^*_{(1440)}$ production, but also for the nucleon
pole contribution, as shown in Fig.\ref{f2}(b). This suggests that
the $\bar pp$ reactions may provide a good place for looking for
those missing $N^*$ resonances with large coupling to $N\sigma$.

For the energy above 2.8 GeV, the contribution from $N^*_{(1680)}$
takes over to be the largest one, produced mainly by t-channel pion
exchange. For each $N^*$ production with t-channel pion exchange,
the contribution from $\bar{N}^*$ is almost four times that from
$N^*$ because of the relevant $C-G$ coefficients for Feynman
diagrams in Figs. \ref{f1}(a) and \ref{f1}(b) except for $N_{(938)}$
where the contribution of the Feynman diagram in Fig. \ref{f1}(c) is
comparable to those from Figs. \ref{f1}(a) and \ref{f1}(b). On the
other hand, the t-channel $\sigma$ exchange cannot produce $\bar
N^*$ to reach the $\bar pn\pi^+$ final state. Therefore, the $N^*$
mainly produced by t-channel pion exchange will show up most clearly
in the $\bar p\pi^+$ invariant mass spectrum, while those $N^*$
mainly produced by t-channel $\sigma$ exchange will show up clearly
only in the $n\pi^+$ invariant mass spectrum.

Here the contribution from $\Delta$ excitation is small in contrast
to the case of the \pp \ reaction, where the $\Delta$ excitation
gives the largest contribution~\cite{ouyang1,ouyangxie}. This is
because the $\Delta^{++}$ excitation in the \pp \ reaction is much
more favored by the isospin $CG$ coefficients than the $\Delta^+$
and $\bar\Delta^0$ excitations in the \ppb \ reaction.

\begin{figure}[htbp] \vspace{-0.cm}
\begin{center}
\includegraphics[width=0.4\columnwidth]{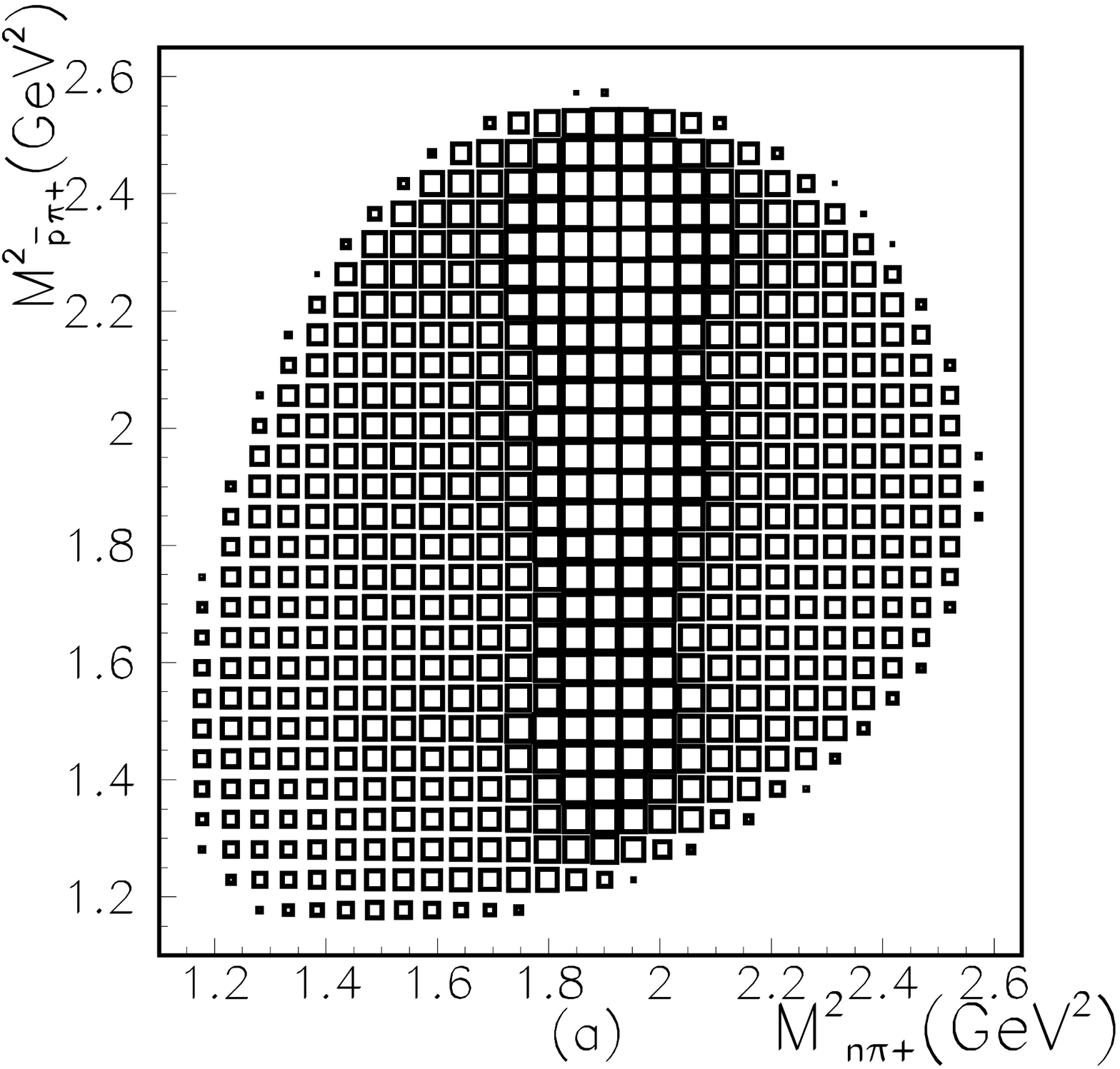}
\includegraphics[width=0.4\columnwidth]{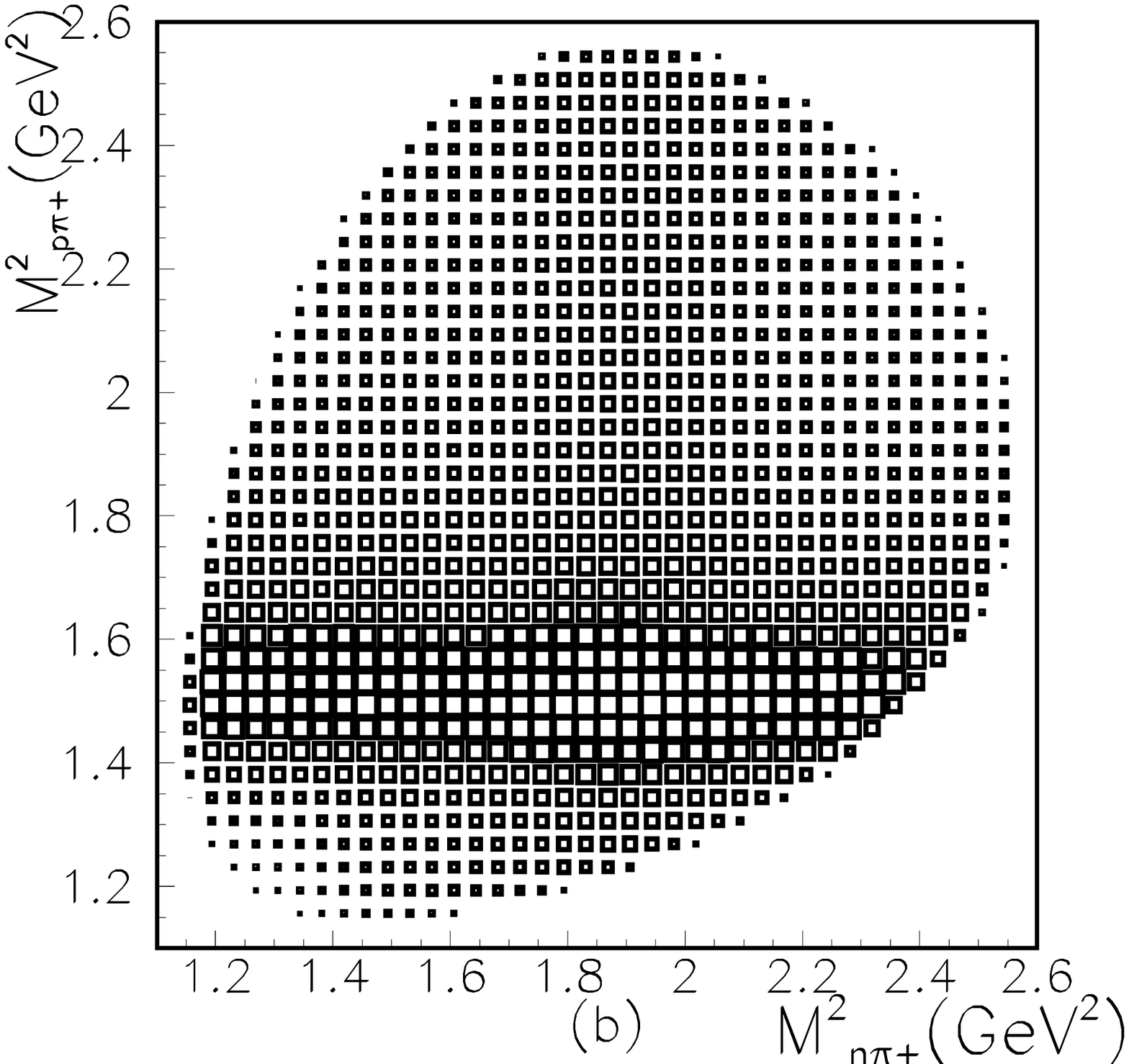}
\includegraphics[width=0.4\columnwidth]{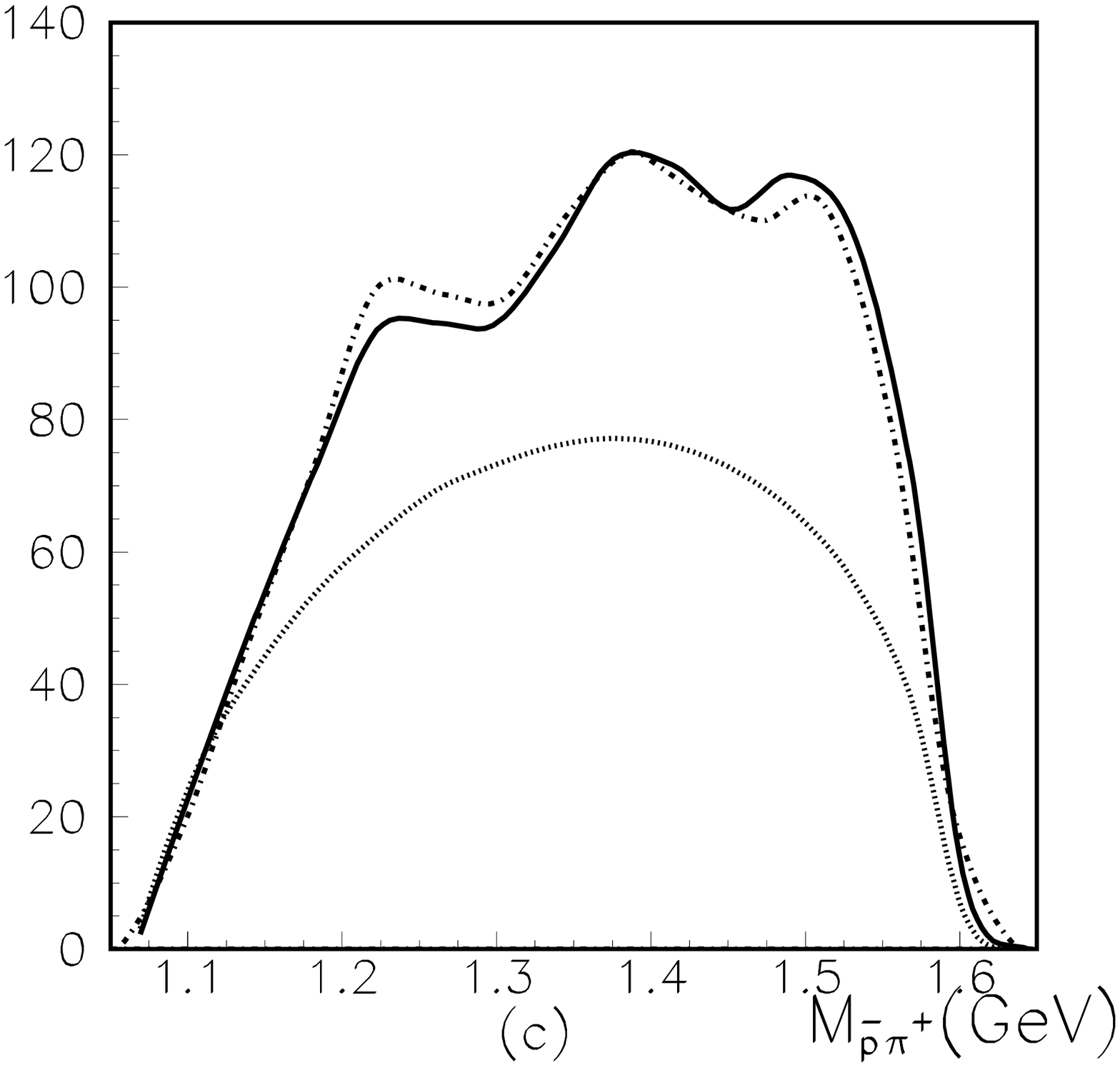}
\includegraphics[width=0.4\columnwidth]{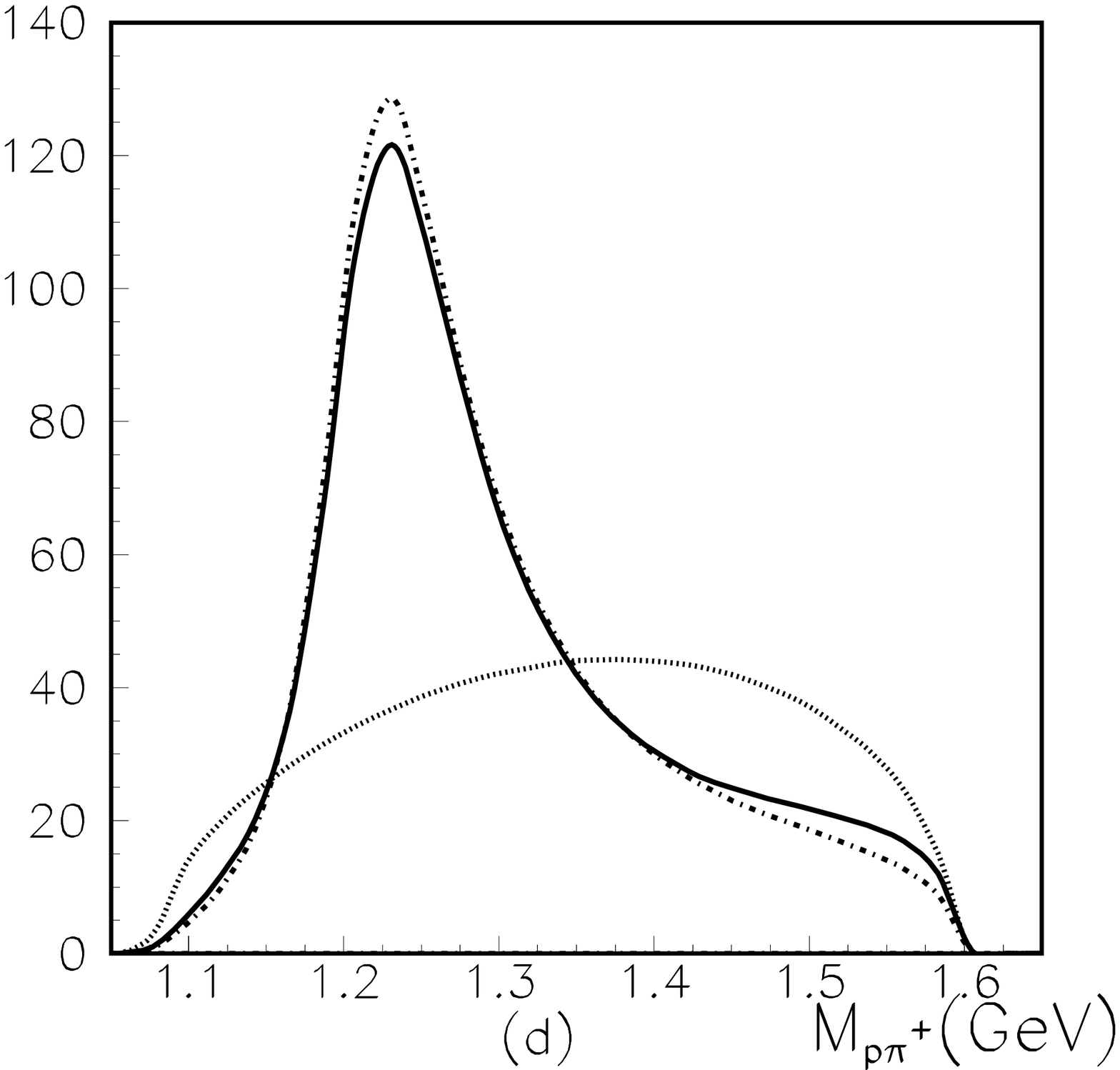}
\includegraphics[width=0.4\columnwidth]{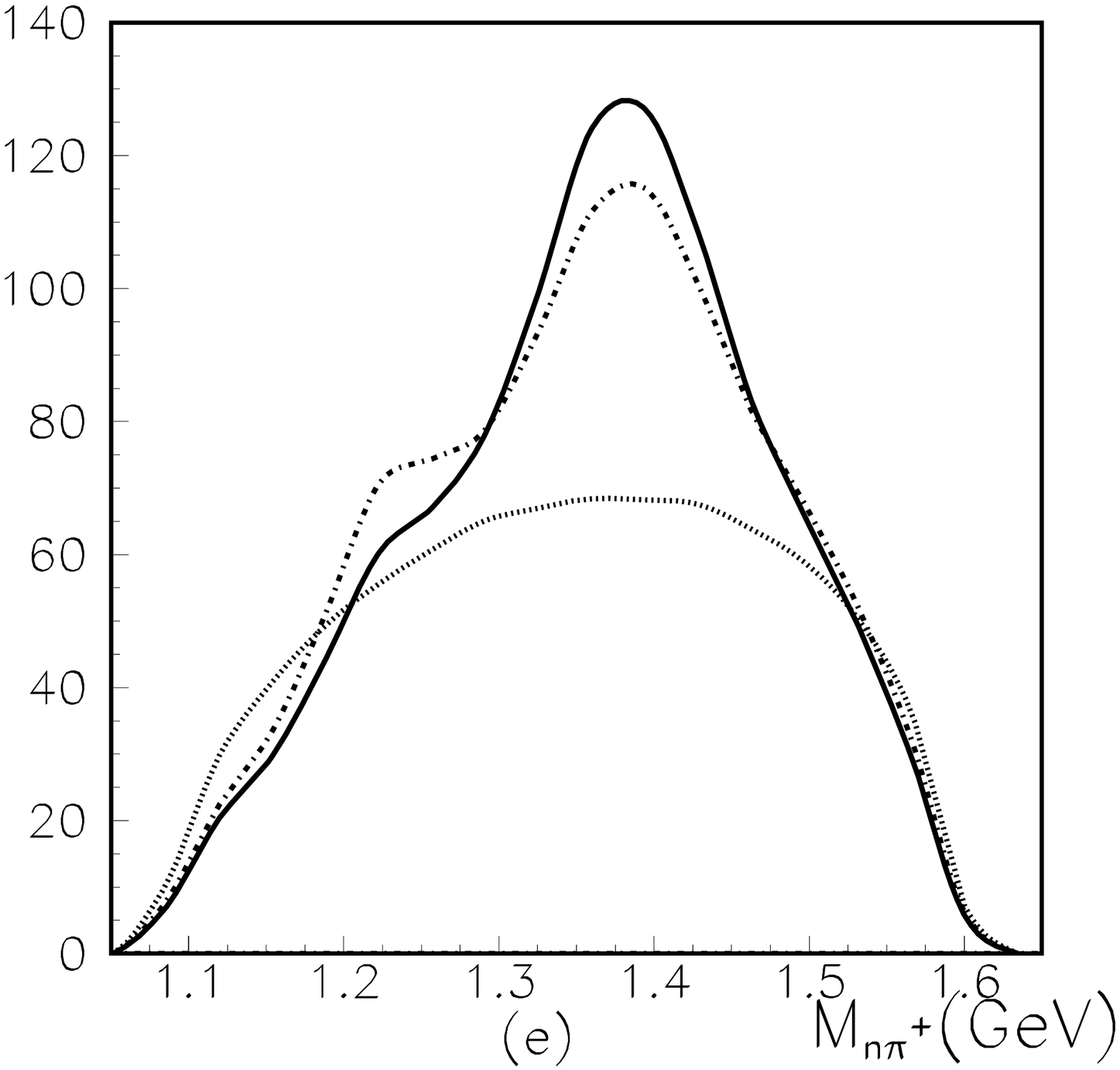}
\includegraphics[width=0.4\columnwidth]{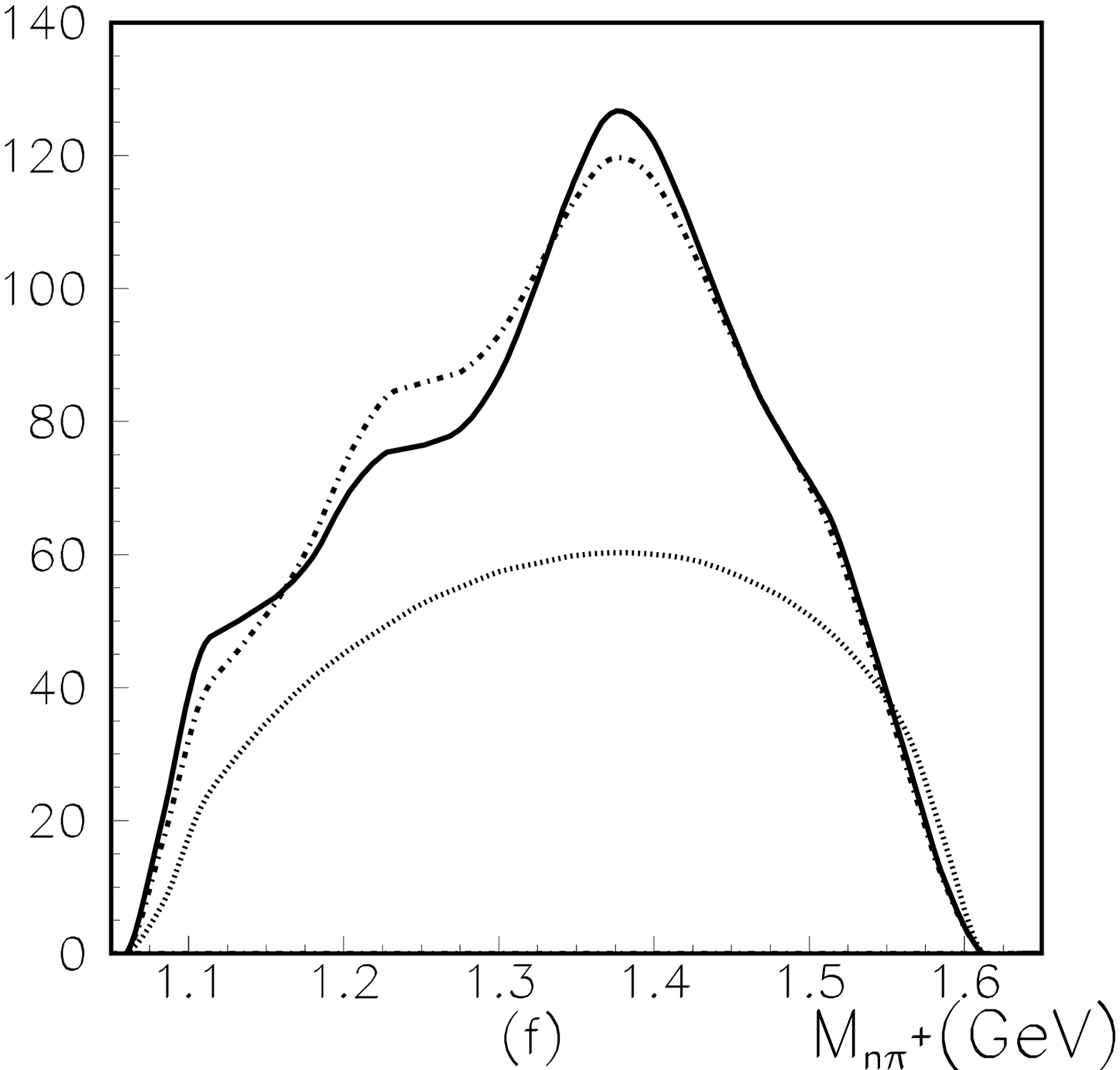}
\caption{Prediction of Dalitz plots and invariant mass spectra
(solid curves) of $\bar{p}\pi^+$ and $n\pi^+$ for the $\bar{p}p \to
\bar{p} n \pi^+$ reaction (left column) compared with the
corresponding ones for the $pp \to p n \pi^+$ reaction (right
column)~\cite{ouyangxie} at $T_{\bar p}=1.55$ GeV. The dotte lines
are results with some parameters replaced by those in
Table.\ref{tab3}. The dashed curves are phase-space distributions.}
\label{fg3}
\end{center}
\end{figure}

\begin{figure}[htbp] \vspace{-0.cm}
\begin{center}
\includegraphics[width=0.4\columnwidth]{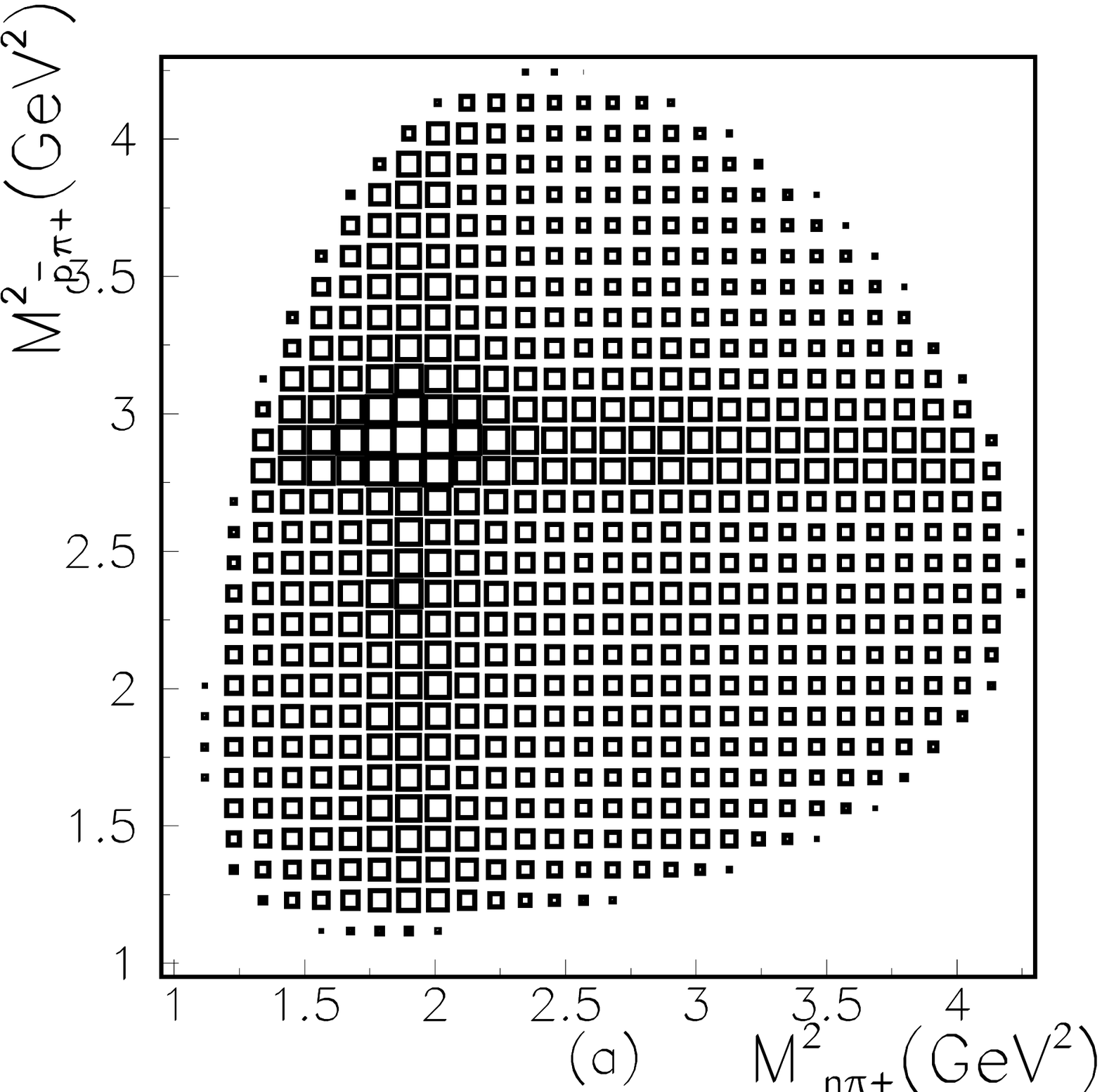}
\includegraphics[width=0.4\columnwidth]{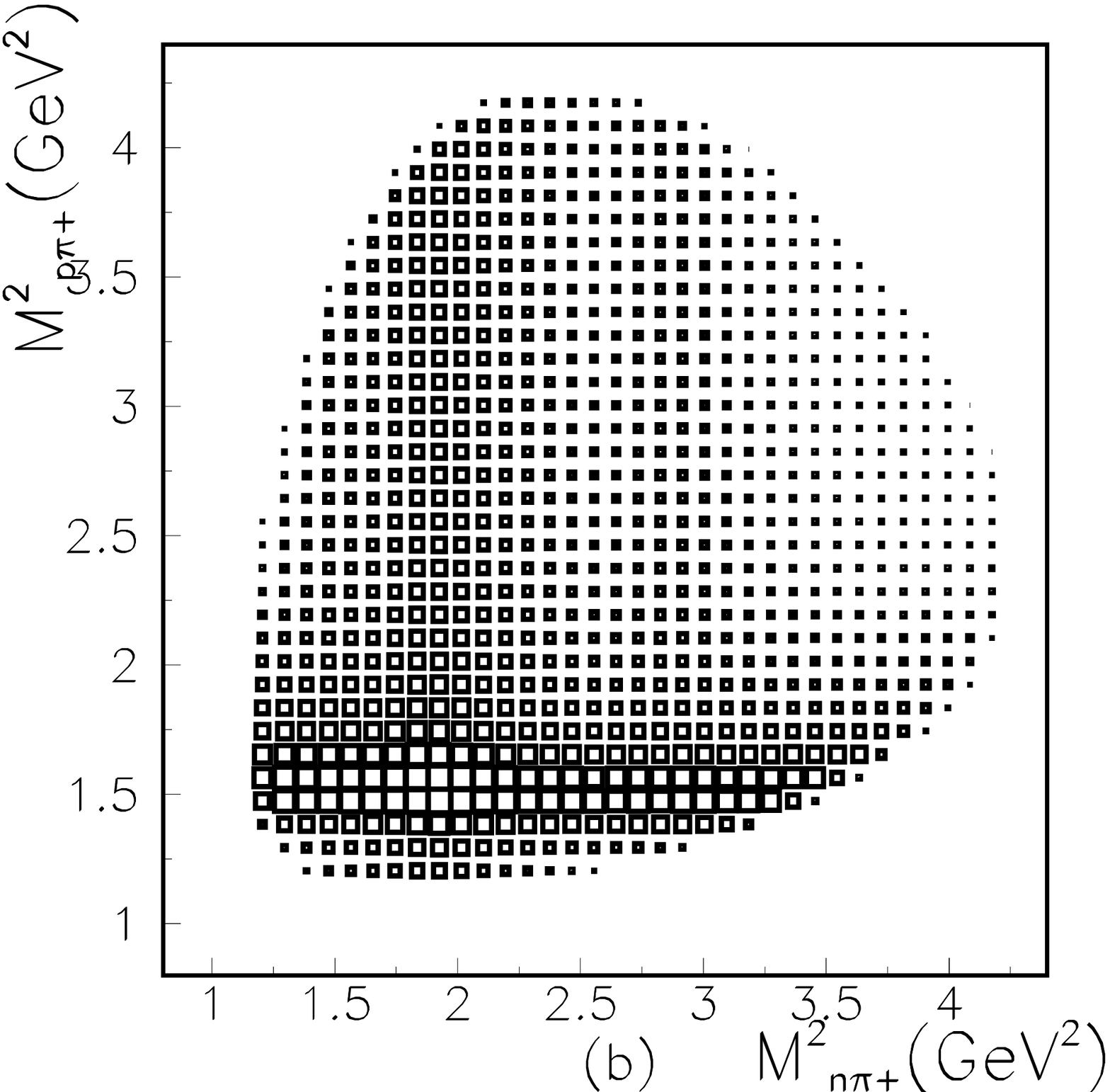}
\includegraphics[width=0.4\columnwidth]{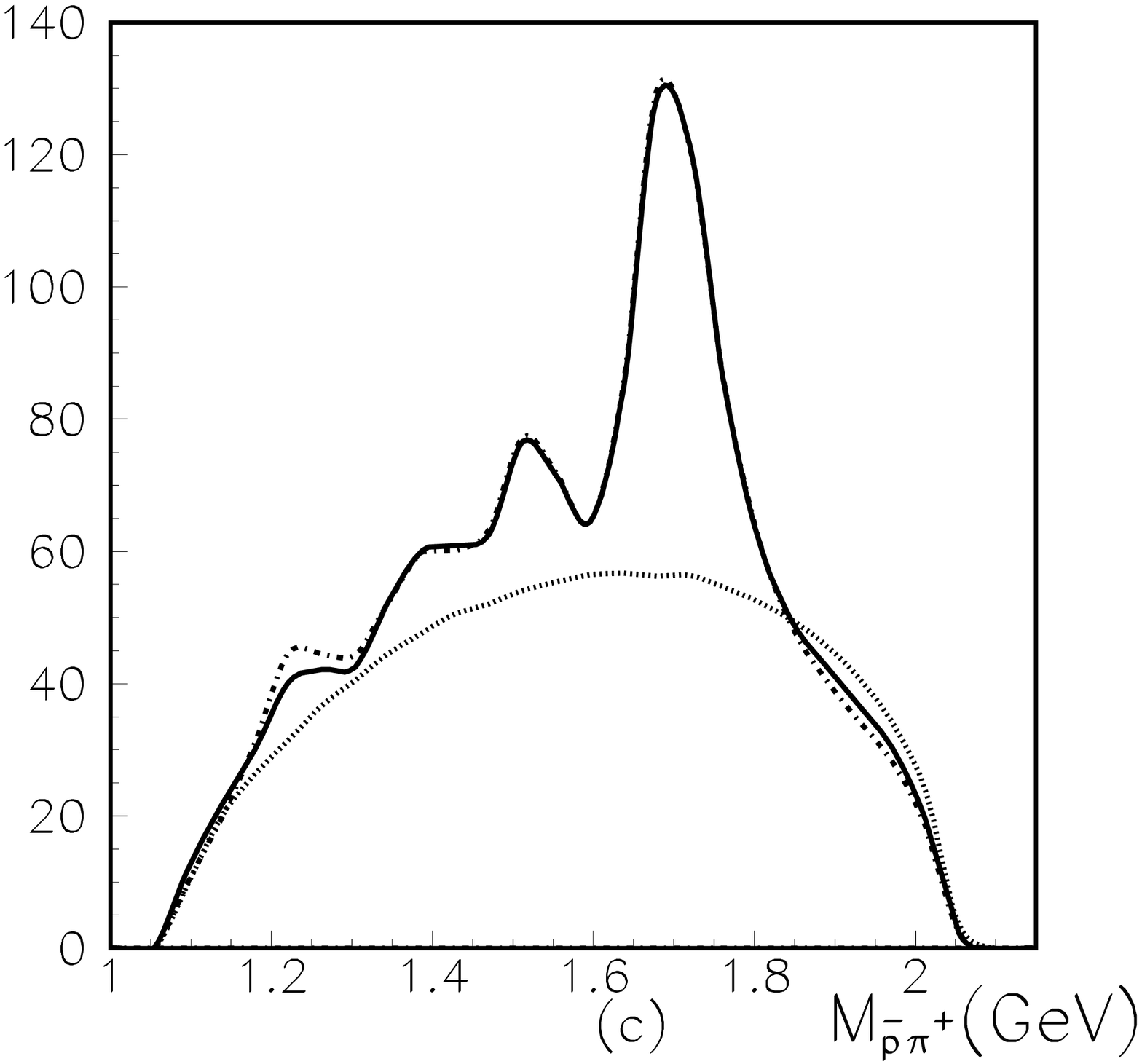}
\includegraphics[width=0.4\columnwidth]{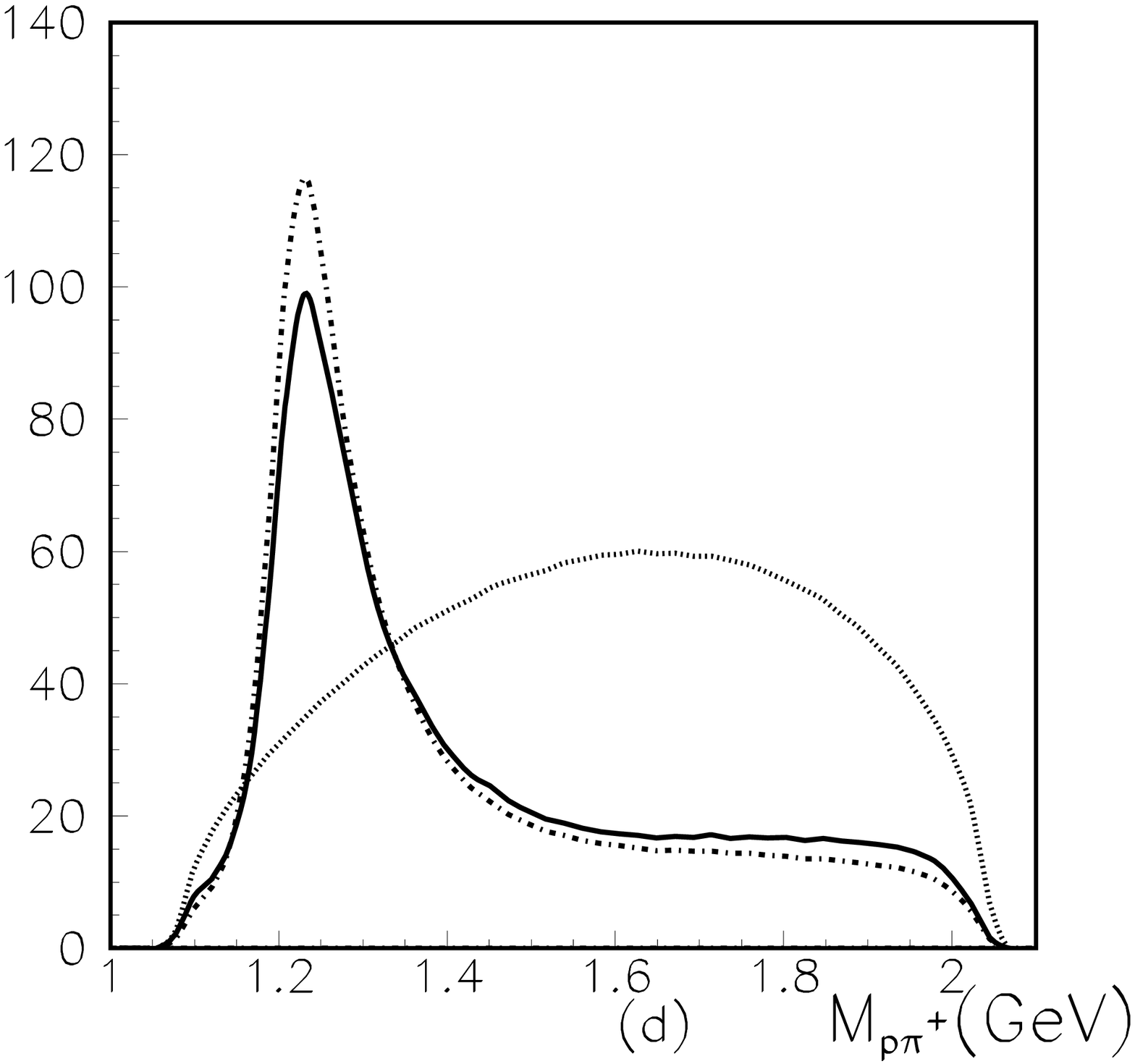}
\includegraphics[width=0.4\columnwidth]{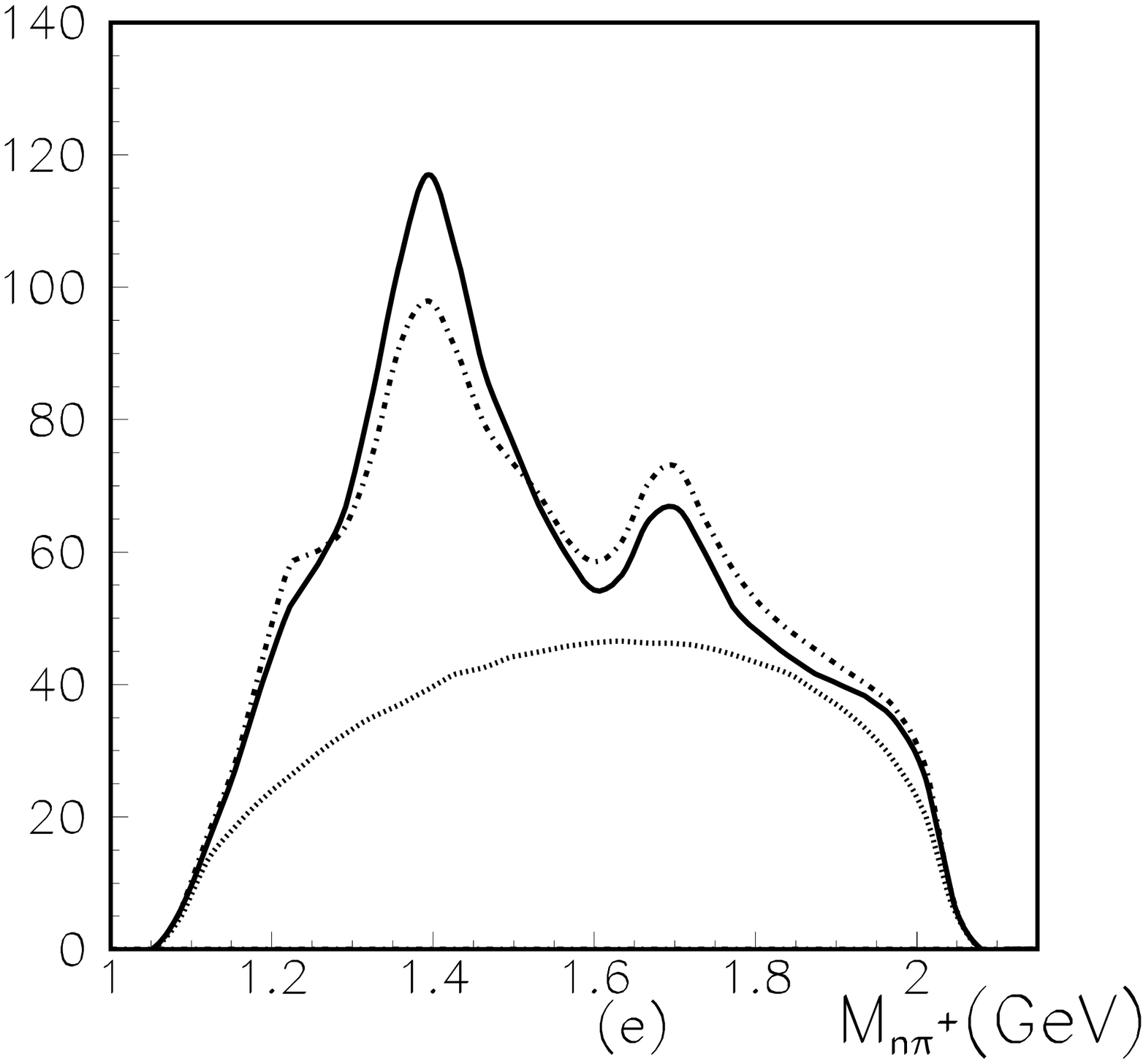}
\includegraphics[width=0.4\columnwidth]{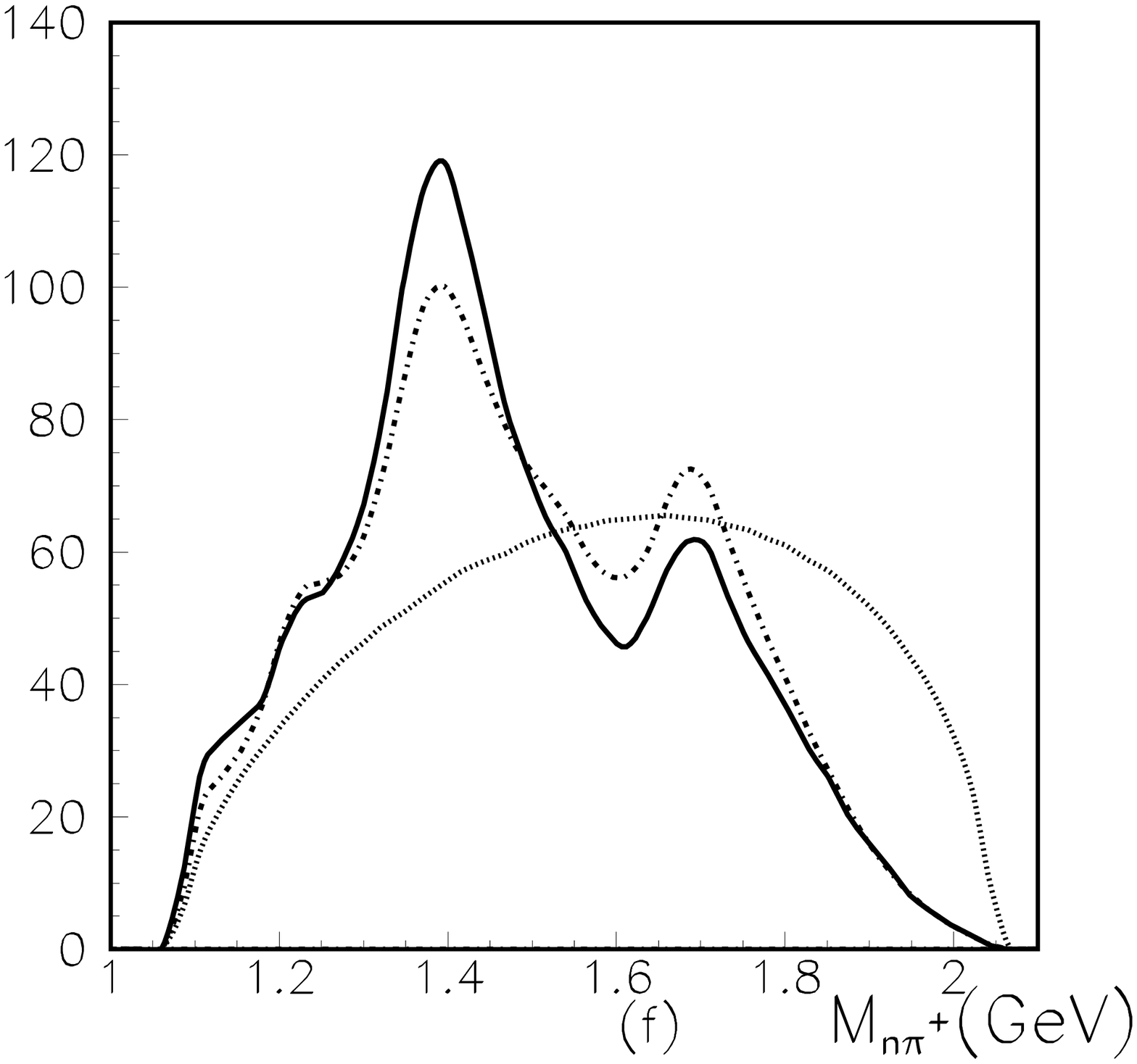}
\caption{Same as Fig. 3, but at $T_{\bar p}=2.88$ GeV.} \label{fg4}
\end{center}
\end{figure}

In Figs.\ref{fg3} and \ref{fg4}, we show the prediction of Dalitz
plots and invariant mass spectra of $\bar{p}\pi^+$ and $n\pi^+$ for
the $\bar{p}p \to \bar{p} n \pi^+$ reaction compared with the
corresponding ones for the $pp \to p n \pi^+$
reaction~\cite{ouyangxie} at $T_{\bar p}=1.55$ GeV (Fig.\ref{fg3})
and 2.88 GeV (Fig.\ref{fg4}).

At $T_{\bar p}=1.55$ GeV, for the \ppb \ reaction, both the Dalitz
plot and $n\pi^+$ invariant mass spectrum show clear dominance of
the $N^*_{(1440)}$ resonance over other contributions. So this
provides us with an excellent place to study the properties of the
Roper resonance. In the $\bar p\pi^+$ invariant mass spectrum, three
peaks correspond to the $\bar\Delta^{0}$, $\bar N^{*0}_{(1440)}$,
and $\bar N^{*0}_{(1520)}$, respectively. In comparison, in the
corresponding $p\pi^+$ invariant mass spectrum for the \pp \
reaction, as shown in Fig.\ref{fg3}(d), one can only see the clearly
dominating $\Delta^{++}$ peak, which shadows all other resonances.
So the \ppb \ reaction here provides also a chance to study some
properties of $ N^{*}(1520)$.

At $T_{\bar p}=2.88$ GeV, in the $\bar p\pi^+$ invariant mass
spectrum, a clear $\bar N^{*0}_{(1680)}$ peak and small $\bar
N^{*0}_{(1520)}$, $\bar N^{*0}_{(1440)}$, and $\bar\Delta^{0}$ peaks
are visible. They are produced by the t-channel pion exchange and
should have their $N^{*+}$ partners making corresponding
contributions to the $n\pi^+$ invariant mass spectrum with a
reduction factor of 4. However, because of the large $N^*_{(1440)}$
production from the t-channel $\sigma$ exchange, the $N^*_{(1440)}$
peak dominates the $n\pi^+$ invariant mass spectrum with a small
$N^*_{(1680)}$ peak in addition. Compared with the \pp \ reaction at
the same energy, the $n\pi^+$ invariant mass spectra are similar,
whereas the $\bar p\pi^+$ spectrum is very different from the
$p\pi^+$ spectrum where the $\Delta^{++}$ peak overwhelmingly
dominates because of its much more favorable isospin factor. For the
\ppb \ reaction, the $\bar N^*$ peaks in the $\bar p\pi^+$ spectrum
put an additional constraint on $N^*$ production from the t-channel
pion exchange. This is an advantage for extracting $N\sigma$
coupling of $N^*$ produced in this reaction.

In our calculation, we have not included the $\bar pp$ initial state
interaction (ISI) and $\bar pn$ final state interaction (FSI)
factors. For the energies considered here, $T_{\bar p}> 1$ GeV,
which is well above the $\bar pp$ threshold, the role of ISI is
basically to reduce the cross section by an overall factor with
little energy dependence~\cite{hanhart1,hanhart3}, and ISI can be
equivalently absorbed into the adjustment of form factor parameters.
This is why the $\Delta N\pi$ form factor that we used is rather
softer than those that include explicitly an additional ISI
reduction factor. Note that the $\bar pp$ elastic scattering cross
sections for the beam energy $T_{\bar p}$ in the range of $1\sim 4$
GeV are larger than the corresponding $pp$ elastic scattering cross
sections~\cite{pdg}. Then the ISI for $\bar pp$ reaction seems not
to give a stronger reduction than the corresponding $pp$ reaction in
this energy range. Assuming the same parameters as for the $pp$
reaction should have given a reasonable estimation of cross sections
for the corresponding $\bar pp$ reaction. For such energies, only a
small portion of $\bar pn$ in the final state will be in the
relative S wave and their FSI should not play a very important role.
Usually, the FSI plays significant role only for near-threshold
meson production.

\begin{figure}[htbp] \vspace{-0.cm}
\begin{center}
\includegraphics[width=0.49\columnwidth]{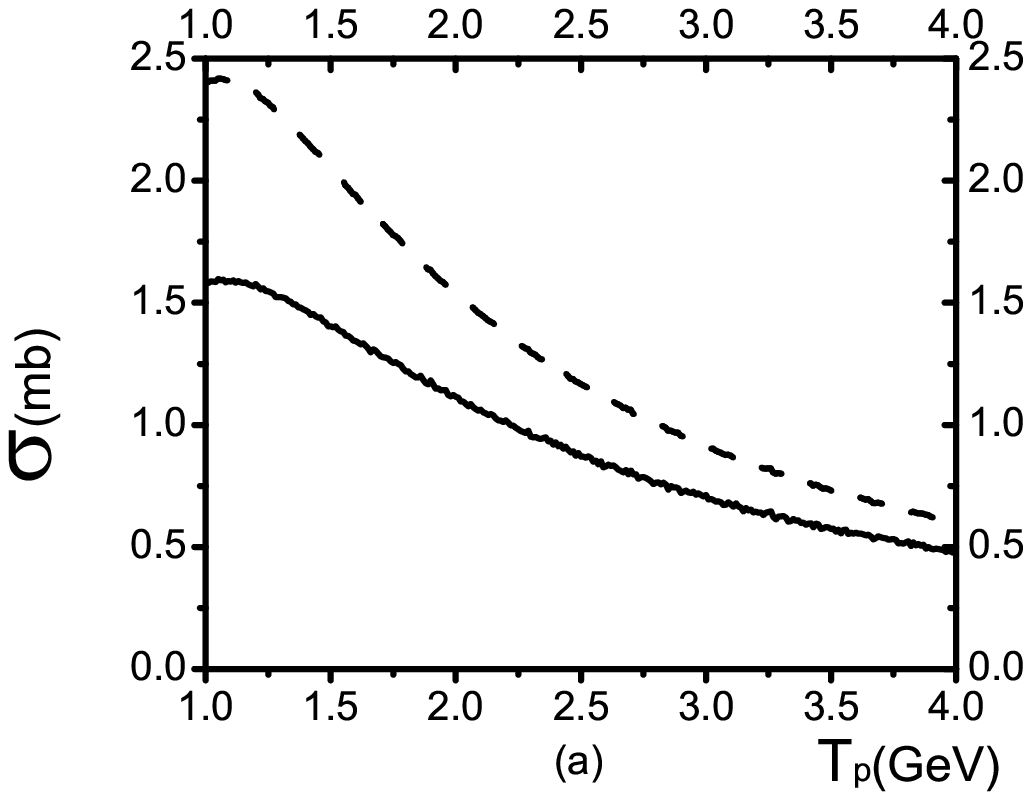}
\includegraphics[width=0.49\columnwidth]{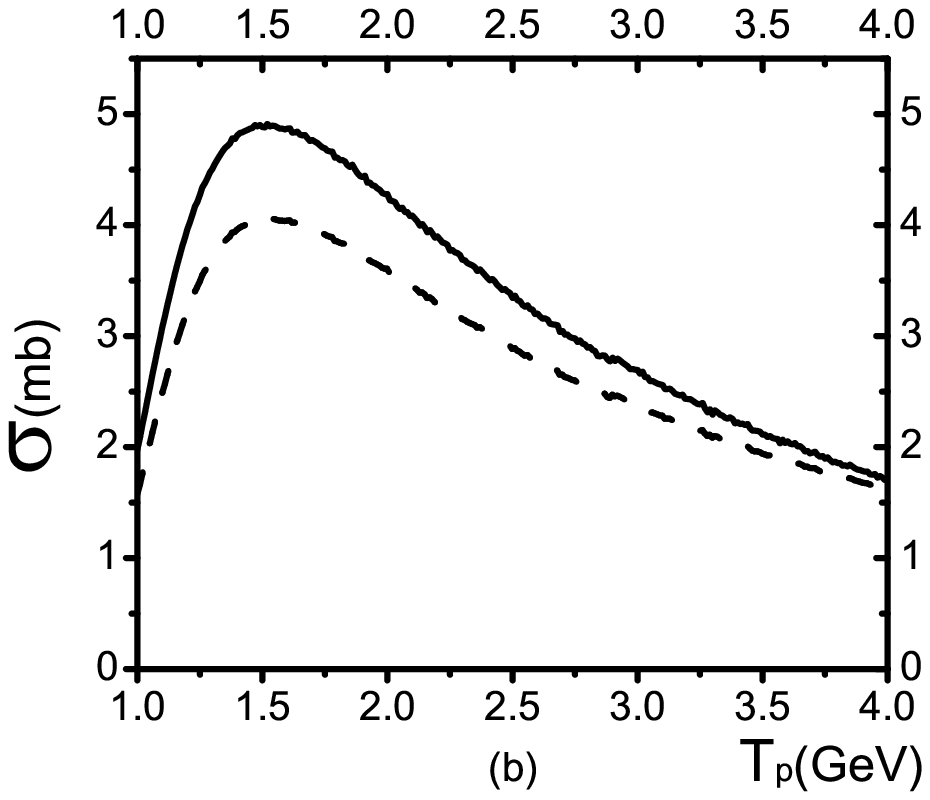}
\caption{Contribution of (a) $\Delta$ and (b) $N^*(1440)$ to the
$\bar{p}p \to \bar{p}n\pi^{+}$ total cross sections with two sets of
parameters: the same parameters as in Ref.\cite{ouyangxie}(solid
curves) and some parameters replaced by those in
Table.\ref{tab3}(dashed curves). } \label{fg5}
\end{center}
\end{figure}

All the above results are based on the parameters in the
Table~\ref{tab1} and \ref{tab2} taken from
Refs.~\cite{ouyang1,ouyangxie}, which reproduce well the data of the
\pp \ reaction~\cite{clement}. For the \pp \ reaction with beam
energies in the range of $1\sim 4$ GeV, the two largest
contributions are found to be from the $\Delta$ excitation with the
t-channel $\pi$ exchange and the $N^*(1440)$ excitation from the
t-channel $\sigma$ exchange~\cite{ouyang1,ouyangxie}. The parameters
for these two biggest contributions were adjusted to reproduce the
data of Ref.\cite{clement}, which demand significant production of
the $N^*(1440)$. But since the data of Ref.\cite{clement} are
preliminary and may not be very reliable~\cite{Skorodko}, the
constraint from the data on the parameters may also be unreliable.
To provide an assessment of the uncertainties involved and their
implications, we should check the results with parameters without
constraint from the data of Ref.\cite{clement}. Some typical values
for these parameters are listed in Table~\ref{tab3}. For the $\Delta
N\pi$ vertex form factor,  the cutoff parameter 1.2 GeV with $n=1$
as in the Bonn model is a commonly used
value~\cite{holinde,rm,jain}. However, with such a hard $\Delta
N\pi$ form factor, an ISI reduction factor about 0.19 is needed to
reproduce the \pp \ total cross section. Without such an ISI factor,
then a much softer $\Delta N\pi$ form factor with cutoff parameter
of about 0.65 GeV is needed to reproduce the data~\cite{kundu}.
According to Ref.\cite{kundu} any attempt to include the
$\rho$-meson exchange worsens the agreement with experiments of
$pp\to n\Delta^{++}$. Hence for the $\Delta$ production, the
$\rho$-meson exchange has been ignored in
Refs.\cite{ouyang1,ouyangxie} and here. Assuming the same parameters
of the Bonn model with the ISI reduction factor of 0.19 for the
$\bar{p}p \to \bar{p}n\pi^{+}$ reaction, then the calculated
contribution of $\Delta$ production to the total cross section is
shown by the dashed curve in Fig.\ref{fg5}(a) and is about 50\%
larger than the result using the parameters of Ref.\cite{ouyangxie}.
For the $N^*_{(1440)}N\sigma$ coupling, the parameters in
Table~\ref{tab3} are from Refs.\cite{hir,oset} which reproduced well
the data on $p\alpha\to p\alpha\pi\pi$ and $pp\to NN\pi\pi$
reactions without including the ISI reduction factor. Assuming these
parameters for the $\bar{p}p \to \bar{p}n\pi^{+}$ reaction, then the
calculated contribution of $N^*_{(1440)}$ production to the total
cross section is shown by the dashed curve in Fig.\ref{fg5}(b) and
is about 30\% smaller than the result using the parameters of
Ref.\cite{ouyangxie}. Note that the $N^*_{(1440)}N\sigma$ coupling
in Table~\ref{tab3} is much smaller than the value in
Table~\ref{tab2}, which is determined from the PDG value for the
$N^*_{(1440)}\to N\sigma$ decay width. So one regards the
$N^*_{(1440)}N\sigma$ coupling in Table~\ref{tab3} as effectively
includeding some ISI reduction factor.

\begin{table}[ht]
\begin{tabular}{ c c c c c}
\hline \hline $R$\ \    & \ $n$\ \    & $g^{2}/4\pi$ & $\Lambda^R_{M}(GeV)$ & Source\\
\hline
 $\Delta_{(1232)}N\pi$    &1    & 19.54       & 1.2 &\cite{holinde,rm}\\
 $N^*_{(1440)}N\sigma$    &1      & 1.33        & 1.7 &\cite{hir,oset}         \\
 $NN\sigma$               &1      & 5.69        & 1.7 &\cite{holinde,rm,oset}         \\
\hline \hline
\end{tabular}
\caption{Parameters without adjustment to fit data from
Ref.\cite{clement}.} \label{tab3}
\end{table}

Results with these parameters without adjustment to fit data from
Ref.\cite{clement} are plotted for various mass spectra of the
$\bar{p}p \to \bar{p}n\pi^{+}$ reaction at $T_{\bar p}=1.55$ and
$2.88$ GeV, as shown by the dashed curves in Figs.\ref{fg3} and
\ref{fg4} as a comparison to those results (solid curves) with the
parameters of Ref.\cite{ouyangxie}. Although quantitatively there
are about 30\%$\sim$50\% uncertainty about the relative production
rates of $\Delta$ and $N^*_{(1440)}$ resonances, qualitatively the
main conclusion of the study is rather firm, {\sl i.e.}, the
$N^*(1440)$ should be clearly seen in the $\bar{p}p \to
\bar{p}n\pi^{+}$ reaction and dominates the reaction at $T_{\bar
p}=1.55$ GeV.

With a clear advantage for studying $N^*$ of the large $N\sigma$
coupling by the \ppb\ reaction, finally let us discuss the
experimental accessibility of this reaction. We know that the
$\bar{p}p$ reaction will be studied by the PANDA (anti-Proton
ANnihilation at DArmstadt) Collaboration at FAIR with the $\bar p$
beam energy in the range of 1.5 to 15 GeV and luminosity of about
$10^{31}cm^{-2}s^{-1}$~\cite{pan1}. For our proposed $N^*$ study
with the \ppb\ reaction, the best beam energy range is 1.5-4 GeV,
with a cross section around 8 mb, which corresponds to an event
production rate of $8 \times 10^{5}$ per second at PANDA/FAIR. The
PANDA is supposed to be a $4\pi$ solid angle detector with good
particle identification for charged particles and photons. For the
\ppb\ reaction, if $\pi^+$ and $\bar p$ are identified, then the
neutron can be easily reconstructed from the missing mass spectrum
against $\pi^+$ and $\bar p$. So this reaction should be easy
accessible at PANDA/FAIR.

In summary, we find that the \ppb\ reaction provides an excellent
place for studying properties of the Roper $N^*(1440)$ resonance and
any other $N^*$ resonances (including some missing ones) with large
couplings to $N\sigma$; and the reaction is easily accessible by the
forthcoming experiments at the PANDA/FAIR. With a large amount of
data on the final states including baryon and antibaryon, the
PANDA/FAIR could play an important role in the study of $N^*$ and
hyperon excited states.

\bigskip
\noindent {\bf Acknowledgements}  We thank Ju-Jun Xie for useful
discussions. This work is partly supported by the National Natural
Science Foundation of China (NSFC) under Grant Nos. 10875133,
10821063, 10635080, the Chinese Academy of Sciences (KJCX3-SYW-N2),
and the Ministry of Science and Technology of China (2009CB825200).

\end{document}